%% file: ORB5.tex
\documentclass[5p]{elsarticle}

\usepackage{subcaption,tabu,ragged2e,tikz,pgfplots,url}
\usetikzlibrary{shapes,3d}

\pgfplotsset{compat=newest}
\pgfplotsset{plot coordinates/math parser=false}
\newlength{\figurewidth}
\newlength{\figureheight}
\pgfplotsset{
  every axis plot post/.append style={
    every mark/.append style={thick, scale=1.5},
  },
  /pgfplots/my xbar legend/.style={
      /pgfplots/legend image code/.code={%
        \draw[##1,/tikz/.cd,bar width=1em]
        plot coordinates { (\pgfplotbarwidth,0.1em)};}
  }
}

\definecolor{mycolor_build_larmor}{HTML}{1F77B4}
\definecolor{mycolor_sort}        {HTML}{E377C2}
\definecolor{mycolor_depos}       {HTML}{2CA02C}
\definecolor{mycolor_solver}      {HTML}{D62728}
\definecolor{mycolor_getfield}    {HTML}{9467BD}
\definecolor{mycolor_gyroavg}     {HTML}{8C564B}
\definecolor{mycolor_push}        {HTML}{FF7F0E}

\begin{document}

\title{Gyrokinetic Simulations on Many- and Multi-core Architectures with the Global Electromagnetic Particle-In-Cell Code ORB5}

\author[epfl]{No\'e Ohana}
\ead{noe.ohana@alumni.epfl.ch}

\author[epfl]{Claudio Gheller}

\author[epfl]{Emmanuel Lanti}

\author[cscs]{Andreas Jocksch}

\author[epfl]{Stephan Brunner}

\author[epfl]{Laurent Villard}

\address[epfl]{Swiss Plasma Center (SPC), \'Ecole Polytechnique F\'ed\'erale de Lausanne (EPFL), Station 13, 1015 Lausanne, Switzerland}

\address[cscs]{Swiss National Supercomputing Centre (CSCS), Via Trevano 131, 6900 Lugano, Switzerland}

\begin{abstract}
Gyrokinetic codes in plasma physics need outstanding computational resources to solve increasingly complex problems, requiring the effective exploitation of cutting-edge HPC architectures. This paper focuses on the enabling of ORB5, a state-of-the-art, first-principles-based gyrokinetic code, on modern parallel hybrid multi-core, multi-GPU systems. ORB5 is a Lagrangian, Particle-In-Cell (PIC), finite element, global, electromagnetic code, originally implementing distributed MPI-based parallelism through domain decomposition and domain cloning.

In order to support multi/many cores devices, the code has been completely refactored. Data structures have been re-designed to ensure efficient memory access, enhancing data locality. Multi-threading has been introduced through OpenMP on the CPU and adopting OpenACC to support GPU acceleration. MPI is further used in combination with the two approaches. The performance results obtained using the full production ORB5 code on the Summit system at ORNL, on Piz Daint at CSCS and on the Marconi system at CINECA are presented, showing the effectiveness and performance portability of the adopted solutions: the same source code version was used to produce all results on all architectures.
\end{abstract}


\maketitle

\section{Introduction}

In magnetized plasmas, the quality of confinement is well known to be degraded by the presence of turbulence arising from various small-scale instabilities, causing cross-field transport to exceed by large factors the neoclassical estimates. This is the main cause of the large size required for a fusion reactor based on the magnetic confinement principle, and consequently of the high cost of such devices. Understanding the fundamental properties of this turbulence is therefore of paramount importance. On the theoretical side, the gyrokinetic theory has been established as the most appropriate and complete description of turbulence in hot, low-collisionality core plasmas. While this dynamical reduction of the full Vlasov-Maxwell system eliminates the fastest scales, solving the resulting gyrokinetic equations remains a problem of formidable complexity requiring sophisticated numerical approaches \cite{Garbet2010}: gyrokinetic equations consist of nonlinear equations for time-dependent distribution functions in 5D phase space coupled to an integro-differential system of time-dependent field equations in 3D. In order to solve increasingly complex problems, outstanding computational resources are required. The effective exploitation of state-of-the-art HPC architectures represents therefore a mandatory task to tackle the most challenging questions in plasma physics.

The pinnacle of modern HPC performance is reached by many-core and heterogeneous computing (based, for instance, on GPU or FPGA devices), and current trends suggest that some form of heterogeneous computing will continue to be prevalent in emerging HPC systems. Therefore, the ability to fully exploit new heterogeneous and many-core architectures is of paramount importance towards achieving optimal performance on modern HPC systems.
On the other hand, with the increasing size and complexity of numerical simulations, it is of primary importance for scientists to be able to exploit all available hardware in emerging HPC environments to achieve maximum computational throughput and efficiency.
Exploiting these various forms of novel hybrid architectures is non-trivial however, due to the challenges presented by mixed hardware computing and the increasing levels of architectural parallelism. New algorithms and numerical and computational solutions are required.

In the last years a massive effort has been devoted to enable scientific applications to hybrid systems in all scientific areas. For plasma physics, we can mention the GPU enabling work made for the GENE code \cite{Dannert2014}, the CGYRO code \cite{Sfiligoi2018}, the GTC code \cite{Meng2013,Madduri2011} and the XGC code \cite{Azevedo2017,Abbott2016}. These works have shown that the effort made at porting gyrokinetic codes to GPU was beneficial: for example, the GTC code was ported on Titan and an acceleration factor up to 3 could be  achieved by using the GPUs \cite{Tang2017}. We note, however, that a thorough benchmark comparison of these gyrokinetic codes is beyond the scope of this paper.

Our target code is ORB5 \cite{Tran1999,Jolliet2007,Bottino2011,Lanti2019}, a state-of-the-art, first-principles-based gyrokinetic code. It is a Lagrangian, delta-$f$, Particle-In-Cell (PIC), finite element, global, electromagnetic code developed by the Swiss Plasma Center (SPC) in collaboration with the Max Planck IPP in Garching and Greifswald and the University of Warwick. Some of its unique features include the use of high order (up to cubic) B-spline basis functions for the field representation, based on the variational Galerkin formulation that allows for the straightforward use of curvilinear toroidal magnetic coordinates including the magnetic axis, flow-conserving source and noise control operators, an enhanced control variate and pullback schemes that solve the cancellation problem for electromagnetic simulations, an adaptive scheme that adjusts the number of Larmor points for the gyroaveraging operator and a field-aligned Discrete Fourier filter eliminating unphysical modes. Last but not least, even though it uses a $\delta f$ representation as control variate, the ORB5 code is equivalent to a truly full-$f$ code, with the exception that the polarization density is linearized.

\sloppy
ORB5 has recently been completely refactored with new data structures and its parallelism enhanced. Particle data structures are now structure of arrays instead of array of structures as it was found to increase the performance by privileging contiguous memory access, both on the CPU and on the GPU. Also, a new data structure was introduced for the representation of points on the Larmor ring that resulted in enhanced modularity and performance. Originally a pure MPI code based on domain decomposition and domain cloning \cite{Kim2000,Hatzky2006}, hybrid MPI/OpenMP and MPI/OpenACC parallel programming models have been introduced. Various multithreading algorithmic solutions have been implemented for the different kernels, in particular for the gyro-averaged charge and current deposition and field assignment.

A remarkable feature of this development is that it has resulted in a single source code version that can be run either on CPU-only or on GPU-equipped HPC systems.
The refactoring has been accomplished adopting a modular approach, making the code easily extensible and maintainable. Continuous integration is supported, through the usage of Git for code versioning \cite{git} and Jenkins for automated testing and verification \cite{jenkins}.

There are other approaches for programming HPC systems in a portable way than the purely directive-based one adopted here. The most popular ones are library based. Kokkos/Cabana is one option which has been used for porting a plasma physics application \cite{Scheinberg2019}. It provides a multidimensional array data structure. A library with a concept similar to Kokkos is Raja \cite{Beckingsale2019}. SYCL is a combination of a library with a specialized compiler \cite{Keryell2019}. We preferred directive-based approach to avoid large parts of our applications becoming library specific.

Being a community-driven code, ORB5 needs to be effectively usable on a large variety of computing systems, so that the scientists can exploit all the available supercomputing resources. Since the development of applications codes such as ORB5 spans a period of time longer than the typical timescale of HPC architecture evolution, our prime objective is to design a portable code which provides good performance on different HPC platforms, rather than fine-tuning the performance on a specific architecture. In this work, we present the results obtained on three cutting-edge supercomputing platforms: Summit at ORNL (Oak Ridge National Laboratory), Marconi at Cineca, and Piz Daint at CSCS (Swiss National Supercomputing Center). The performed tests are representative of typical production runs, and only the compilation procedure accounts for specific, architecture-related optimization, adaptable by any standard user. Too specific architecture-related customization has been avoided, privileging portability and usability instead of extreme performance optimization.  All results in this paper have been obtained with a single source code version which is actually the full production code version. The numerical and physical parameters of the data sets correspond to real production cases. Most importantly, all timings reported include all host-device and all across-node data transfers.

The details of the ORB5 code and its parallel implementation will be given in Section \ref{sec:orb5}, followed, in Section \ref{sec:results}, by the presentation of the results obtained for different tests and benchmarks. Results will be discussed in Section \ref{sec:conclusion} together with the main conclusions.

\section{The ORB5 code}
\label{sec:orb5}

ORB5 is a global gyrokinetic PIC code solving for the Vlasov-Maxwell set of equations in Tokamak geometry. Its model is derived from a variational formulation with consistent ordering \cite{Sugama2000,Tronko2017}. It accounts for electromagnetic perturbations around a realistic MHD equilibrium \cite{Lutjens1996} or an approximated ``ad-hoc'' one. Different options are available for the field solver, namely the long wavelength approximation, Pad\'e approximation, or arbitrary wavelength \cite{Dominski2017}. One can simulate multiple gyrokinetic species, and drift-kinetic, adiabatic or hybrid electrons. Among others, the code also features inter- and intra- species collisions \cite{Vernay2010}, heat sources \cite{McMillan2008}, and strong flows \cite{McMillan2011}. A detailed description of ORB5 capabilities can be found in \cite{Lanti2019}.

\subsection{Algorithms and MPI implementation}
\label{sec:numeric}
In ORB5 a control variate scheme is used, meaning that it represents the distribution functions of plasma species as the sum of a time-independent function $f_0$ and a time-dependent perturbation $\delta f$. The latter is discretized with numerical markers (also called particles in this paper). Those markers evolve in fields discretized with 3D finite element B-splines up to third order. ORB5 uses the magnetic coordinates $(s,\theta^\star,\varphi)$, where $s$ is the radial coordinate, $\varphi$ the toroidal coordinate, and $\theta^\star$ the poloidal coordinate modified from the geometric poloidal angle so that field lines are straight in this coordinate system. The inner radial boundary condition can be Dirichlet, `free' or unicity if the magnetic axis is part of the domain. The outer boundary condition is Dirichlet. The matrix problems resulting from the 3D finite element discretization of field equations are solved in discrete Fourier space in the poloidal and toroidal dimensions, allowing to take advantage of the anisotropy of the fluctuations to filter out unphysical perturbation modes which are not aligned with the field lines. This Fourier filtering reduces the required number of markers of one order of magnitude for the same noise level \cite{Jolliet2009}. Moreover, thanks to the background and profiles axisymmetry, toroidal mode numbers are decoupled from each other in the field solvers, offering a trivial level of parallelism. In order to control the noise inherent to PIC simulations, ORB5 includes a Krook operator, a coarse graining technique, and quadtree smoothing. The electromagnetic cancellation problem in Amp\`ere's law can be cured using an enhanced control variate scheme \cite{Mishchenko2017} or the pullback scheme \cite{Mishchenko2019}. The gyro-averaging operations make use of numerical markers along the Larmor rings of each guiding center, as shown in Figure \ref{fig:larmor}. The number of Larmor points per guiding center can be fixed or adaptive (scaling with the ring size).

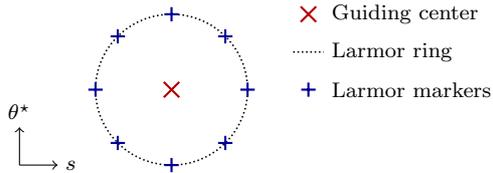
\begin{figure} \centering \footnotesize
\begin{tikzpicture}[scale=1]
    \draw[->] (-2,-1) -- (-1.5,-1);
    \node[anchor=west] at (-1.5,-1) {$s$};
    \draw[->] (-2,-1) -- (-2,-.5);
    \node[anchor=south] at (-2,-.5) {$\theta^\star$};
    \draw (0,0) node[cross out, thick, draw=black!30!red, inner sep=0pt, minimum size=5pt] {};
    \draw[densely dotted, semithick](0,0) circle (1);
    \foreach \i in {1,...,8}{
        \draw ({cos(45*\i)},{sin(45*\i)}) node[cross out, rotate=45, thick, draw=black!40!blue, inner sep=0pt, minimum size=3pt] {};
    }
    \draw (1.8,1) node[cross out, thick, draw=black!30!red, inner sep=0pt, minimum size=5pt] {};
    \node[anchor=west] at (2,1) {Guiding center};
    \draw[densely dotted, semithick](1.6,.5) -- (2,.5);
    \node[anchor=west] at (2,.5) {Larmor ring};
    \draw (1.8,0) node[cross out, rotate=45, thick, draw=black!40!blue, inner sep=0pt, minimum size=3pt] {};
    \node[anchor=west] at (2,0) {Larmor markers};
\end{tikzpicture}
\caption{Larmor markers around a guiding center in the poloidal plane. The number of Larmor markers per guiding center can be fixed or proportional to the ring size.}
\label{fig:larmor}
\end{figure}
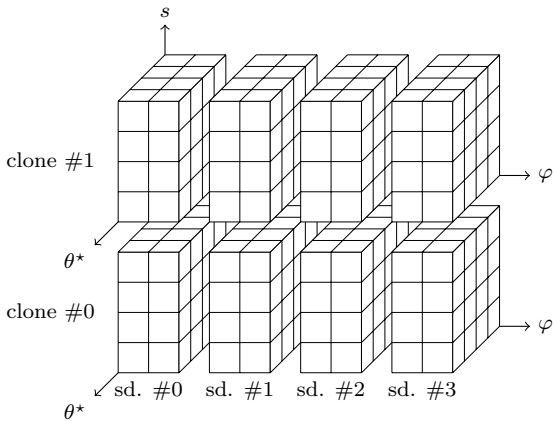
\begin{figure} \centering \footnotesize
\begin{tikzpicture}[scale=.4, black]
    \draw[->] (0,0,-4) -- (0,10,-4);
    \node[anchor=south] at (0,10,-4) {$s$};
    \draw[->] (0,0,-4) -- (12,0,-4);
    \node[anchor=west] at (12,0,-4) {$\varphi$};
    \draw[->] (0,5,-4) -- (12,5,-4);
    \node[anchor=west] at (12,5,-4) {$\varphi$};
    \draw[->] (0,0,-4) -- (0,0,2);
    \node[anchor=north east] at (0,5,2) {$\theta^\star$};
    \draw[->] (0,5,-4) -- (0,5,2);
    \node[anchor=north east] at (0,0,2) {$\theta^\star$};
    \foreach \iclone in {0,...,1}{
        \foreach \isd in {0,...,3}{
            \begin{scope}[canvas is xy plane at z=0]
                \fill[white] (3*\isd,5*\iclone) rectangle (3*\isd+2,5*\iclone+4);
                \draw[very thin] (3*\isd,5*\iclone) grid (3*\isd+2,5*\iclone+4);
            \end{scope}
            \begin{scope}[canvas is yz plane at x=3*\isd+2]
                \fill[white] (5*\iclone,0) rectangle (5*\iclone+4,-4);
                \draw[very thin] (5*\iclone,0) grid (5*\iclone+4,-4);
            \end{scope}
            \begin{scope}[canvas is xz plane at y=5*\iclone+4]
                \fill[white] (3*\isd,0) rectangle (3*\isd+2,-4);
                \draw[very thin] (3*\isd,0) grid (3*\isd+2,-4);
            \end{scope}
        }
        \node[anchor=east] at (-0.5,2+5*\iclone) {clone \#\iclone};
    }
    \foreach \isd in {0,...,3}{
        \node[anchor=north] at (1+3*\isd,0) {sd. \#\isd};
    }
\end{tikzpicture}
\caption{MPI 2-level parallelization of ORB5 in subdomains (sd.) and clones. Toroidal dimension can be decomposed into subdomains, and each of those subdomains can be cloned to split the particle workload over many MPI tasks.}
\label{fig:parallelization}
\end{figure}
ORB5 has two levels of MPI parallelism (Figure \ref{fig:parallelization}). The first one is domain decomposition along the toroidal dimension. Each subdomain works with field and particle data of a toroidal slice. This slicing is uniform in order to guarantee a balanced workload. Hence, the number of subdomains is constrained to be a divider of the number of toroidal grid cells, which is frequently a power of 2 for best Fast Fourier Transform performance. There are three kinds of communications between subdomains. First, they send and receive particle data leaving or entering the subdomain. Particles can jump across several subdomains in a time step so it is an any-to-any communication. Due to the axisymetry of the underlying physical problem, the number of markers per subdomain is guaranteed to be statistically uniform at all times. Second, subdomains communicate guard cell field data with their neighbors due to the finite element overlap over several grid cells. Third, parallel data transposition is required before performing local (to a CPU) Fourier transform in the toroidal direction. This operation is an all-to-all communication which swaps the partitioned dimension between toroidal and poloidal ones.

The second level of MPI parallelization in ORB5 is domain cloning. It consists in replicating subdomain field data on different tasks, thus splitting the particle workload without changing the grid resolution. MPI communications among clones are reductions of the grid data before carrying out the field solvers, and broadcasts afterwards. Domain cloning was introduced since the very first version of the ORB5 code. It allowed for running with more MPI tasks than the number of toroidal grid points. In the newly refactored, multithreaded version of ORB5 (see below), domain cloning is shown to be useful on multi-socket nodes, i.e. the optimum configuration is to use one clone per socket.

Instead of (or in addition to) domain cloning, we may consider 2D or 3D domain decomposition: the potential scalability gain of these options was investigated in Ref.\cite{Jocksch2017}. While 1D decomposition in the toroidal direction poses no problem for load balancing, additional decomposition in the radial and poloidal directions is more challenging: ensuring particle load balance requires equal-volume domains, but due to the polar-like grid, this leads to off-balance grids. Moreover, the finite Larmor radius (FLR), lying essentially in the poloidal plane, introduces technical difficulties when a Larmor ring straddles two or more domains: this requires additional guard cells but unlike those required by the overlap of finite elements, the number of necessary guard cells depends on the largest Larmor radius of all markers in the domain and is not necessarily the same throughout the plasma, nor is it the same in all simulations. Although subject of investigation, the 2D or 3D domain decomposition has not yet been implemented in the ORB5 code.

\begin{figure} \centering \footnotesize
\tikzstyle{linesep}=[thick, dashed, gray]
\tikzstyle{point}=[circle, radius=.1, fill=mainthemecolor]
\tikzstyle{arrow}=[->, >=latex, thick, black!80!white]
\tikzstyle{mytext}=[draw, ellipse, inner sep=1pt, text=white, semithick]
\newcommand{\GC}
{\begin{tikzpicture}[scale=.1, black!30!red]
\draw[fill](1,1.5) circle (.3);
\draw[fill](2,3) circle (.3);
\draw[fill](6,1.5) circle (.3);
\draw[fill](5,5) circle (.3);
\draw[fill](3,3.6) circle (.3);
\draw[fill](1.2,5.5) circle (.3);
\end{tikzpicture}}
\newcommand{\larmor}
{\begin{tikzpicture}[scale=.1, black!40!blue]
\draw[thin](1,1.5) circle (1);
\draw[thin](2,3) circle (2);
\draw[thin](6,1.5) circle (1.3);
\draw[thin](5,5) circle (1.5);
\draw[thin](3,3.6) circle (0.7);
\draw[thin](1.2,5.5) circle (1.2);
\end{tikzpicture}}
\newcommand{\field}
{\begin{tikzpicture}[scale=.1, black!60!green]
  \begin{scope}[canvas is xy plane at z=0]
    \draw[very thin] (0,0) grid (4,4);
  \end{scope}
  \begin{scope}[canvas is yz plane at x=4]
    \draw[very thin] (0,0) grid (4,-4);
  \end{scope}
  \begin{scope}[canvas is xz plane at y=4]
    \draw[very thin] (0,0) grid (4,-4);
  \end{scope}
\end{tikzpicture} }
\begin{tikzpicture}[xscale=1.5,yscale=1.4]
\draw[linesep] (2,-1) -- (2,2.5);
\draw[linesep] (4,-1) -- (4,2.5);
\node[black!30!red] at (1,2.2) {Guiding center data};
\node[black!40!blue] at (3,2.2) {Larmor ring data};
\node[black!60!green] at (5,2.2) {Field data};
\node (A) at (1,1.2) {\GC};
\node (B) at (3,1.2) {\larmor};
\node (C) at (5,1.2) {\field};
\node (D) at (5,0) {\field};
\node (E) at (3,0) {\larmor};
\node (F) at (1,0) {\GC};
\draw[arrow, out=42, in=138] (A) to node[mytext, fill=mycolor_build_larmor]{\strut build Larmor} (B);
\draw[arrow, out=42, in=138] (B) to node[mytext, fill=mycolor_depos]{\strut deposit} (C);
\draw[arrow, out=-48, in=48] (C) to node[mytext, fill=mycolor_solver]{\strut solve} (D);
\draw[arrow, out=-138, in=-42] (D) to node[mytext, fill=mycolor_getfield]{\strut get fields} (E);
\draw[arrow, out=-138, in=-42] (E) to node[mytext, fill=mycolor_gyroavg]{\strut gyroaverage} (F);
\draw[arrow, out=132, in=-132] (F) to node[mytext, fill=mycolor_push]{\strut push} (A);
\end{tikzpicture}
\caption{Main stages of an ORB5 time step. The 6-fold scheme differs from the usual PIC 4-fold scheme because of the introduction of the Larmor ring data.}
\label{fig:main_stages}
\end{figure}
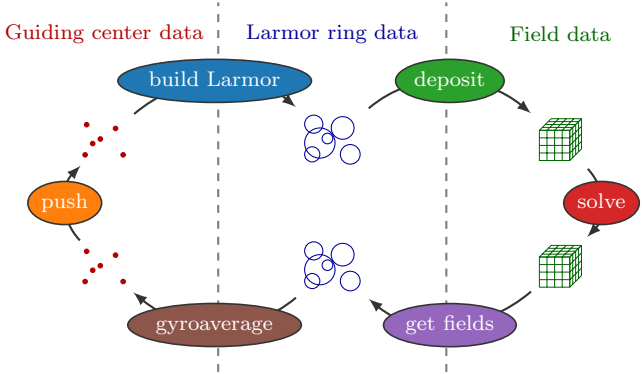
The time integrator is a fourth-order Runge-Kutta.

\subsection{Rewriting and refactoring strategy} \label{sec:refactoring}
Developers of legacy codes are confronted to the recurrent dilemma: should the code be entirely rewritten, or is refactoring a better option? We have adopted a two-pronged approach that combines both aspects of rewriting and refactoring. First, we have extracted the fundamental kernels of the original ORB5 code and rewrote them from scratch. These then served as `test-bed' codes, which we called `PIC-engine' \cite{Hariri2016} and `GK-engine' \cite{Ohana2016} for the pure PIC and for the specific gyrokinetics aspects, respectively. With respect to ORB5, the GK-engine misses the toroidal geometry, electromagnetic fluctuations, noise controls, heat sources, and multiple kinetic species, among other features. Meanwhile, the PIC-engine does not even include any field solver, nor gyrokinetic effects. In these test-beds, several options for data structures and multithreading were implemented and tested. These include in particular the challenging particle-to-grid and grid-to-particle kernels. Then, the most performing options of the PIC-engine and GK-engine were implemented into the full production ORB5 code.

A standard PIC time step consists of 4 stages: (1) the deposition of particle charges and currents on the grid; (2) the field solve on the grid; (3) the interpolation of the field at particle positions; and (4) the particle push.
As compared to standard PIC schemes, the finite Larmor radius (FLR) of the gyrokinetic description introduces an additional non-locality.
In the original ORB5, data structures were attached to guiding centers and stages (1) and (3) were treated as follows: a loop on guiding centers, inside of which a loop on Larmor markers was performed. As the number of Larmor markers can be different for every guiding center, this nested loop combination performed weakly and proved difficult to optimize. Moreover, trying to improve data locality by sorting according to guiding center positions does not ensure locality of Larmor points.

Therefore, another strategy was adopted for the refactoring of the ORB5 code: a new data structure was introduced, named `Larmor ring data', containing the positions in real space of the Larmor markers and guiding center weights.
The main stages of an integration substep now consist of 6 stages instead of 4 and are depicted in Figure \ref{fig:main_stages}.
With this, the operations related to FLR (`build Larmor' and `gyroaveraging') can be completely separated from those of particle-to/from-grid (`deposit' and `get fields') for which now all optimizations pertinent to `standard' (non-gyrokinetic) PIC algorithms can then be straightforwardly applied, including improving data locality by sorting the Larmor point according to their position in the grid.
This strategy was first tested in the GK-engine and then, since it proved beneficial, was introduced in the refactored ORB5 code. The introduction of an intermediate Larmor data structure makes the code more efficient as well as more modular, at the expense of a higher memory footprint. Indeed, loops over Larmor points are taken out of the loops over guiding centers, leading to better vectorization. It also allows for the sorting of the Larmor markers by cells before depositing them, improving data locality and avoiding random memory accesses. Indeed, without sorting, the deposition routine loops over the markers and then over the finite elements they contribute to, while with sorting the outermost loop can be on the grid cells and the innermost one on the particles in that cell.


\subsection{Multi/Many threads acceleration}

Multi-threading for multi-core CPUs has been implemented by using the OpenMP application programming interface \cite{OpenMP}. The current implementation follows version 3.1 standard. OpenMP represents a consolidated and stable solution to exploit shared memory devices, ensuring good scalability up to many cores and high portability, being supported by all the major hardware and software providers and implemented in the most common compilers of, in particular, Fortran, C and C++ programming languages. It is based on a set of directives which instruct the code on how to split the work among the available threads, with minimum impact on the source code, although specific customization has to be done in order to ensure good performance and scalability.

In ORB5, OpenMP is used exclusively in routines looping over the particles. Such loop iterations are in general independent from one another, which make them trivially parallelizable over OpenMP threads. When it comes to accumulation of some marker quantity on a grid, such as charge and current depositions or binning diagnostics, one has to take care about race conditions (several threads updating the same memory location simultaneously). For scalar quantities, we use the OpenMP built-in \texttt{reduction} clause. For arrays, we make each OpenMP thread accumulate to a temporary replica, and we reduce them afterwards. This manual operation was found to be more efficient than the \texttt{reduction} clause, especially for large arrays. Those techniques have been found to be much more efficient than \texttt{atomic} directives. The deposition algorithms using particle sorting avoid this grid replication by multithreading the outer loop over grid cells so that the inner loop over markers within a cell can be executed serially.

The GPU implementation has followed a similar approach, adopting the OpenACC programming model \cite{OpenACC} to enable the GPU to carry out the most time-consuming parts of the code. The current implementation reads revision 2.5. Just like OpenMP, OpenACC is a directive-based approach, which allows keeping the source code almost unchanged, leaving to the compiler the task of creating the GPU version. Contrary to OpenMP however, OpenACC is a new standard under continuous development and only few compilers, namely PGI and GNU, fully support its latest releases. Hence portability is still limited in this respect. On the HPC systems typically targeted by ORB5 (and, in particular, used in this work), the PGI compiler is expected to be available. The adoption of OpenACC is therefore not limiting its usability.

OpenACC provides two main classes of directives: those to manage work sharing and those to control data transfer. Both are designed with performance as the main objective.
The OpenACC approach was tested on the simplified PIC-engine kernels \cite{Hariri2016} and several options regarding the data layout and multithreading were tried there, the most performant being implemented in ORB5 afterwards.

In ORB5, all the computations involving particles have been offloaded to the GPU: more precisely, these are the stages `push', `build Larmor', `deposit', `get fields' and `gyroaverage', with only the `solve' operations performed on the host. Loops over particles are particularly well suited for GPUs since millions of independent iterations can be executed by thousands of threads. Communication in between GPU and CPU is minimized by keeping particle data on the GPU only. More precisely, all Larmor ring data are kept strictly on the GPU, while only a fraction of the guiding center data, related to the markers that change domain, has to be transferred via MPI at every time step from GPU to GPU.
Having a single kernel using particles on the CPU would annihilate the speed-up brought by the GPU due to the time required to transfer particle data between device and host.

The `solve' stage consists of a computational part, namely the backsolve operations of the field solver and the MPI communication of grid data required by the toroidal domain decomposition and toroidal discrete Fourier transform, in addition to reductions of grid data over the clones.
Field backsolve operations have not yet been ported to GPUs as they were initially not found to be the critical parts for our application.
Since particle-to/from-grid operations are performed on the GPU, field data has to sit on both device and host. Therefore field data is transferred from the GPU to the CPU after the deposition, and back to the GPU after the field solve.

In the current ORB5 implementation, most of the loops multithreaded with OpenMP and OpenACC are the same. Therefore, when enabling OpenACC to benefit from GPUs, OpenMP is disabled. The situation could evolve in the future, for instance if the `solver' stage gets multithreaded on the host with OpenMP.
One can also think of getting rid of either OpenMP or OpenACC directives through the adoption of more recent standards. Indeed, OpenMP 4.5 provides GPU instrumentation directives, and OpenACC 2.7 enables one to treat CPUs as devices. Those standards were not properly implemented by the compilers at the time of this work.

Concerning race conditions in deposition routines, the grid replication method used with OpenMP threads is not suitable anymore with GPU threads due to their large number. We therefore use OpenACC \texttt{atomic} directives around accumulation instructions in order to make them thread safe. The introduction of OpenACC \texttt{atomic}s for our application on NVIDIA GPUs, where the atomic update has been effectively implemented at hardware level on Pascal and Volta architectures, does not lead to any significant time overhead compared to racy depositions. It is to be noted that these race conditions can be totally avoided by sorting particles as mentioned in the previous section (\ref{sec:refactoring}) and multithreading the loop over grid cells.

\section{Results}
\label{sec:results}

\begin{table*}[t]
\centering
\footnotesize
\tabcolsep 5pt
\renewcommand{\arraystretch}{1.6}
\begin{tabu} to \textwidth {l X[3] l >{\RaggedRight}X[1.5] X[1.1] >{\RaggedRight}X[3]}
System    & Computing Node & $N_\mathrm{nodes}$ & Interconnect & Compiler & Compiler flags\\
\hline
Marconi   & 2 Intel Xeon 8160 CPU (Skylake)   & 3224 & Fat-Tree Intel Omnipath & Intel 17.0.4.196 & \texttt{-O2 -xCORE-AVX2 -fopenmp}\\
Piz Daint & Intel Xeon E5-2690 v3 (Haswell) CPU + NVIDIA Tesla P100 GPU& 5704 & Dragonfly Aries & PGI 18.10\newline + Cuda 9.1 & \texttt{-O3 -fast -Minline -Munroll -Mvect=levels:7,nosizelimit -Mcuda -acc -ta=tesla,cc60}\\
Summit    & 2 IBM POWER9 CPUs \mbox{+ 6 NVIDIA} Tesla V100 GPUs & 4608 & Fat-Tree EDR 100G InfiniBand & PGI 18.10\newline + Cuda 9.2 & \texttt{-O3 -fast -Minline -Munroll -Mvect=simd [-mp=nonuma | -Mcuda -acc -ta=tesla,cc70]}\\
\end{tabu}
\caption{Technical characteristics of the three adopted HPC platforms, the adopted compilers and the corresponding compilation options. Two sets of options have been used on Summit, enabling either OpenMP or OpenACC exclusively.}
\label{tab:system}
\end{table*}
As stated above, all the results presented in this paper have been obtained with a single source code version, which is the full production ORB5 code version. Also, the simulation parameters used in this paper are representative of production runs. Most importantly, all timings reported include all host-device and node-to-node data transfers.

In order to analyze the performance and the scalability of our parallel, multi-threaded implementation of ORB5, we have performed a number of tests and benchmarks exploiting three different state-of-the-art architectures available at Cineca, CSCS and ORNL supercomputing centers.
Cineca is the Italian national HPC facility. We have used the A3 partition of its {\it Marconi} flagship system, made of 3224 2$\times$24-cores Skylake CPU nodes. We have exploited the GPU partition of {\it Piz Daint} at CSCS, the Swiss National Supercomputing Center. This partition is a Cray XC50 machine, with 5704 nodes equipped each with a 12-cores Intel Haswell CPU and one NVIDIA Tesla P100 GPU. Finally, we have run on the {\it Summit} supercomputer at Oak Ridge National Laboratory in the USA, deploying 4608 nodes, each equipped with two 21-cores IBM POWER9 CPUs and 6 NVIDIA Tesla V100 GPUs. Table \ref{tab:system} summarizes the characteristics of the systems and indicates the version of the adopted compilers as well as invoked compiler flags.

As physical and numerical simulation parameters we use gyrokinetic deuterium ions with an adaptive number of Larmor points as well as ``heavy'' kinetic electrons only 200 times lighter than ions. The pullback scheme is used to solve for the electromagnetic fluctuations. The Poisson equation makes use of the long wavelength approximation. A density-, zonal flow- and parallel flow- conserving Krook operator is used as noise control and heat source. The number of time steps has been set to be a small fraction of an actual simulation.

The single node tests of section \ref{sec:node}, as well as those of sections \ref{sec:arch} and \ref{sec:sort} adopt a mesh of $N_s\times N_{\theta^\star}\times N_\varphi=512\times1024\times4$ cells and $20\cdot10^6$ particles for each of the two simulated species (deuterium and electrons). This setup is representative of one subdomain, typically fitting on a single node, of a medium-size tokamak production run. Meshes and particle numbers relevant for production runs on multiple nodes will be used in section \ref{sec:scaling}.

For all the tests, we have adopted as performance metric the time-to-solution and the fractions of time spent in the different parts of the code. Only time spent in the time loop has been considered, thus excluding initialization. In a typical production run there are at least tens of thousands of time steps, so that the initialization time is a negligible fraction of the total.

\subsection{Single node MPI and OpenMP performance}
\label{sec:node}

ORB5 allows specifying different configurations of parallelism exploiting the domain decomposition, cloning and multithreading (on the CPU or on the GPU). A first set of tests has focused on the CPU implementation, analyzing how different configurations influence the performance and selecting Skylake as the reference architecture (the comparison among different architectures will be given in the next section).

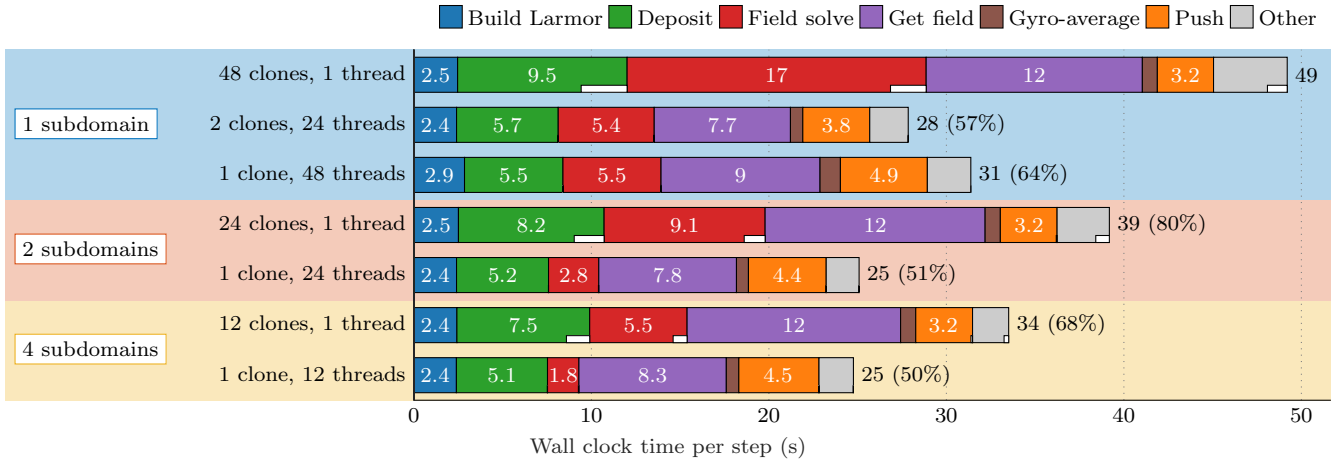
\begin{figure*}%
  \centering%
  \setlength{\figurewidth}{.95\linewidth}%
  \setlength{\figureheight}{.2667\figurewidth}%
  \footnotesize%
  \input{figures/MPI_vs_OpenMP.tikz}%
  \vspace{-3mm}%
  \caption{Comparison of timings for different MPI and OpenMP parallelization combinations on a Skylake node (2$\times$24-cores). Timing `Other' includes in particular sources and diagnostics. The `Field solve' includes discrete Fourier transforms and backsolve operations. White bars represent time spent in MPI routines. The percentages of the total time are relative to the top bar.}%
  \label{fig:MPI_vs_OpenMP}%
\end{figure*}

Figure \ref{fig:MPI_vs_OpenMP} shows the breakdown of the time to solution in the different components of the code, as introduced in section \ref{sec:numeric}. Each bar represents a different combination of number of MPI subdomains (domains in which the grid is partitioned along the toroidal direction), number of MPI clones (replicas of a domain), and number of OpenMP threads per MPI task. The product of numbers of subdomains, clones and threads equals the number of CPU cores (48 in this case).

The first three bars show the timings for one MPI subdomain, the second two bars for 2 subdomains, and the last two bars for 4 subdomains. The first bar of each group uses pure MPI parallelization, without multithreading. The number of OpenMP threads is then increased, decreasing the number of MPI clones accordingly.

With 1 subdomain, going from 48 clones to 2 clones times 24 threads leads to an overall speed-up of 43\% (intermediate steps are not shown, but the performance is monotonically improving with the number of threads). Shared memory, besides optimizing the memory usage by avoiding data replica, improves the performance compared to clones, which have to account also for the overhead of MPI reduce and broadcast operations. When running with OpenMP, the threads of a task are bound to contiguous cores. Going to full OpenMP parallelization with 48 threads means that they will expand over the two sockets of the Skylake CPU. Such configuration performs worse than keeping one MPI task per socket, because threads access memory resources on the remote socket.

Although not yet itself OpenMP parallel, the field solver is also faster when threads are adopted for the other parts of the code. In that case, fewer clones are instanced, resulting in a more efficient memory usage. Memory contention between several clones, concurrently solving for their field replica, has in fact proved to lead to a significant slow-down of the calculation.

Charge and current deposition is much faster with OpenMP threads than MPI clones, especially as a result of improved reduction.
The `Get field' routine is faster as well because field is interpolated from a single instance per socket in memory, not from many clones.
The push loop slow-down with OpenMP threads is due to the embedded calculation of diagnostics within the particle loop which leads to false sharing, with thread replicas allocated in the same cache line. This will be fixed in the coming release of the code, leading to a further improvement of the overall performance with respect to pure MPI setups.

Skylake's capability to run two threads on a single core (so-called hyperthreading) was not found to bring any significant speed-up to the overall timing of the code. It was not the case on other architectures either. The results are not shown here, but some parts of the code were accelerated by a small fraction of the order of 10\%, while others were slowed down by a similar fraction. Getting a significant speed-up would require fine-tuning the hyperthreading on the different kernels of the code, which we decided not to do in order to stay consistent with our portability policy.

With 2 subdomains, the pure MPI parallelization with 24 clones is 20\% faster than with 1 subdomain. In particular, the field solver timing is reduced almost by a factor 2, the mesh size per domain being halved. The deposition is a bit faster as well because data locality is improved by having sorted the particles in toroidal domains 
and the MPI reduction over clones is alleviated. Trading MPI clones for OpenMP threads leads to an extra 36\% speed-up for the same reasons as the 1-subdomain case.

The best performance is obtained with 4 subdomains, with a 32\% speed-up on the pure MPI parallelization compared to the 48 clones case, and a 50\% speed-up with OpenMP. The number of subdomains cannot be increased further for this test case with $N_\varphi=4$ because it has reached the limit of a single toroidal grid cell per subdomain.
Alleviating this limit would require 2D or 3D domain decomposition \cite{Jocksch2017} but this option is not implemented in ORB5 yet.

\subsection{Running on different architectures}
\label{sec:arch}

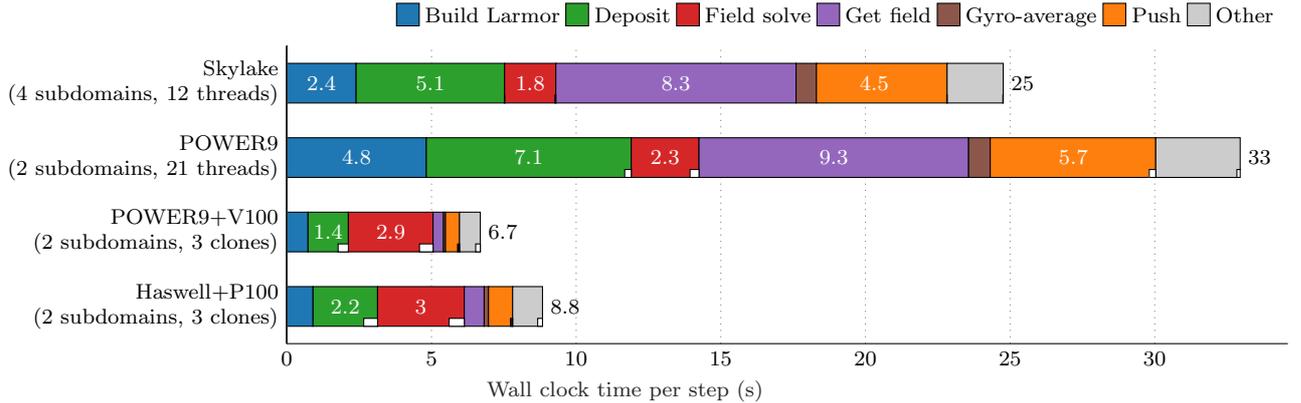
\begin{figure*}%
  \centering%
  \setlength{\figurewidth}{.95\linewidth}%
  \setlength{\figureheight}{.2265\figurewidth}%
  \footnotesize%
  \input{figures/architecture_comparison.tikz}%
  \vspace{-3mm}%
  \caption{ORB5 performance on different architectures: Skylake 2$\times$24-cores CPU, POWER9 2$\times$21-cores CPU with and without 6 V100 GPUs, and 6 Haswell 12-cores CPUs with one P100 GPU each. Timing `Other' includes in particular sources and diagnostics. The `Field solve' includes discrete Fourier transforms and backsolve operations. White bars represent time spent in MPI routines.}%
  \label{fig:architecture_comparison}%
\end{figure*}

The different available computing architectures have been exploited, selecting for each of these the best possible combination of number of subdomains, clones and threads, fully exploiting the flexibility given by the various levels of parallelism implemented in ORB5, taking into account a number of constraints of the code (e.g. the number of grid points in the toroidal direction must be divisible by the number of subdomains) and specific node features in terms of number of cores and GPUs. These tests have been performed with the aim of comparing how different architectures perform in practice for our application code. On the other hand, exactly the same source code was used, i.e. there was no fine tuning, architecture-dependent optimization.
For all GPU-equipped architectures, one GPU is used per MPI rank.
We did not investigate other options of e.g. OpenMP threads linked to different GPUs, multiple MPI ranks oversubscribing the same GPU, or one MPI task directing multiple GPUs, which might bring further optimization.
The resulting configurations are summarized in Figure \ref{fig:architecture_comparison}, where the breakdown of the timings is presented. The first two bars show the results on the Skylake and the POWER9 CPUs, using MPI and OpenMP. The fraction of time spent in the various kernels is approximately the same for both CPUs.
The difference in performance measured for all the code components between the two architectures can be partially justified with the larger number of cores available on Skylake and with the usage of the PGI compiler on the POWER9, which may provide sub-optimal optimization compared to the native IBM XLF compiler. Further investigation is required in order to fully understand this difference.

The third and fourth bar of Figure \ref{fig:architecture_comparison} show the performance when GPUs are used. Due to the memory constraints, the test requires at least 4 GPUs to run. We chose to fully exploit the Summit node with its 6 GPUs and, consistently, 6 Piz Daint nodes. Two subdomains times three clones were set in both cases. OpenMP multithreading was switched off, since our current implementation of ORB5 does not support the simultaneous usage of OpenACC and OpenMP.
On both systems, the field solver is the most time-consuming kernel. This is in fact the only part of the code which has no GPU implementation, running on the CPU only. For this field solver kernel, as seen in Section \ref{sec:node}, the usage of the clones penalizes the POWER9+GPU setup compared to the POWER9 one, due to the presence in the former of collective MPI operations, necessary to coordinate the work on the different clones. All the other kernels run on the GPU. Comparing the P100 and V100 GPUs, we find an overall performance difference of a factor 1.5 (subtracting the time spent on the field solver using the CPU only), which results from the combination of the faster computing capabilities and the higher memory bandwidth of the V100 architecture.

The analysis of the runtime of ORB5 on a Summit node running with or without GPUs, show that the usage of the accelerators for this application reduces the computing time by a factor of $\sim$5 overall, and by a factor of $\sim$8 when only GPU-enabled kernels are considered.
The kernel benefiting most from the GPU acceleration is `get field' with a factor $\sim$30 speed-up.
The push routine is accelerated by a factor $\sim$12, the gyro-averaging by $\sim$10, the Larmor construction by $\sim$7 and the deposition by $\sim$5. Let us recall that these timings include the CPU-GPU data transfers and that, for reference, the peak performance ratio of GPU acceleration on a Summit node, in the absence of any data transfer, is $\sim$40.

\subsection{Particle sorting}
\label{sec:sort}

\begin{figure*}%
  \centering%
  \setlength{\figurewidth}{.9\linewidth}%
  \setlength{\figureheight}{.1833\figurewidth}%
  \footnotesize%
  \input{figures/sorting.tikz}%
  \vspace{-3mm}%
  \caption{Comparison with and without sorting on Skylake (4 subdomains times 12 threads) and Power+V100 architectures (2 subdomains times 3 clones). Timing `Other' includes in particular sources and diagnostics. The `Field solve' includes discrete Fourier transforms and backsolve operations. White bars represent time spent in MPI routines.}%
  \label{fig:sorting}%
\end{figure*}
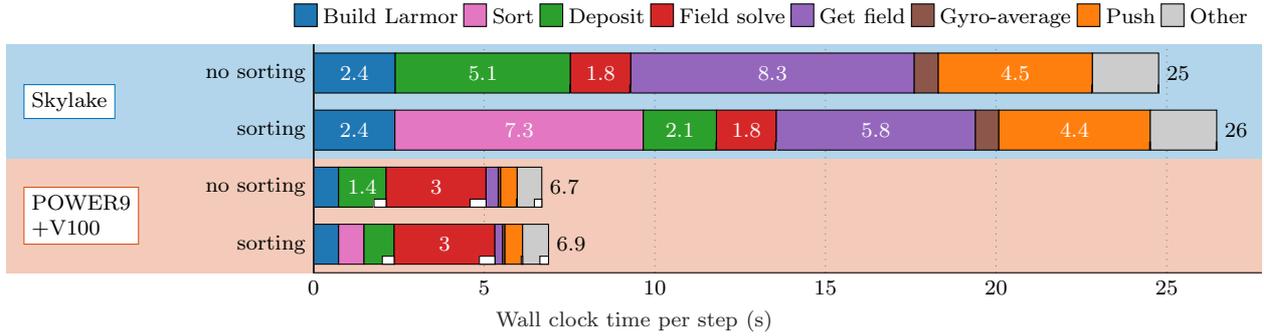

Data locality is expected to be highly beneficial for multi-core, multi-thread applications. For this reason, as mentioned in Section \ref{sec:numeric}, sorting of markers with respect to the field cells has been introduced with the ORB5 refactoring.
A bucket sort algorithm is adopted, this choice resulting from the analysis performed in \cite{Jocksch2015}, where several algorithmic variations and GPU implementations were investigated. One should however note that the results of the cited publication were performed on GPUs using Kepler architecture, which do not implement hardware \texttt{atomic} operations in double precision.

Figure \ref{fig:sorting} shows the effect of sorting on Marconi (Skylake) and Summit (POWER9+V100). On Marconi, the deposition routine is 59\% faster with sorting than without and `get field' is 31\% faster. Unfortunately, the sorting itself becomes the most time-consuming part of the code (28\% of the overall time) and counterbalances the speed-up it brought. In the end, the code is 7\% slower with sorting for the present simulation parameters.

Even though the deposition and `get field' timings are not the dominant ones on Summit with the GPUs, particle sorting has been tried and found to improve deposition by 66\% (ignoring MPI reduction time) and `get field' by 36\%. The overall performance is however again decreased by 3\% due to the cost of sorting itself.

Those results demonstrate the potential of the adoption of particle sorting to enforce memory locality, which however can be fully exploited only if specific optimization of the sorting kernel is accomplished. Indeed, only the most promising sorting algorithm found in the PIC-engine studies \cite{Jocksch2015} has been implemented in ORB5 but it may not be the best for ORB5.

\subsection{Parallel scalability}
\label{sec:scaling}

\begin{figure*}
  \setlength{\figurewidth}{.49\linewidth}
  \setlength{\figureheight}{.618\figurewidth}
  \footnotesize
  \raggedright 
  \begin{subfigure}{.48\linewidth}
    \centering
    \input{figures/scalability_marconi.tikz}\\
    \subcaption{Marconi, 2 clones times 24 threads per node. Speed-up is normalized to the smallest case on a single node, which takes 20s per time step.}%
    \label{fig:scalability_marconi}
  \end{subfigure}
  \\\vspace{5mm}
  \centering
  \begin{subfigure}{.48\linewidth}
    \centering
    \input{figures/scalability_daint.tikz}\\
    \subcaption{Piz Daint, 12 threads per node for CPU-only runs, and a single task per node for CPU+GPU runs. Speed-up is normalized to the smallest case on a single node without GPU, which takes 115s per time step.}
    \label{fig:scalability_daint}
  \end{subfigure}
  \hfill
  \begin{subfigure}{.48\linewidth}
    \centering
    \input{figures/scalability_summit.tikz}\\
    \subcaption{Summit, 2 clones times 21 threads per node for CPU-only runs, and 6 clones per node for CPU+GPU runs. Speed-up is normalized to the smallest case on a single node without GPU, which takes 28s per time step.}
    \label{fig:scalability_summit}
  \end{subfigure}
\caption{Strong and weak scalings on Marconi (a), Piz Daint (b) and Summit (c),  supercomputers. Each series represents a strong scaling. The set of first points from each strong scaling provides a weak scaling, multiplying the number of nodes and the problem size by a factor 8 for both particles and grids between consecutive colors. Continuous lines with open symbols indicate results using CPUs only, while dashed lines with filled symbols make use of the GPUs. We use as many subdomains as number of nodes up to the limit of the number of toroidal grid cells, beyond which clones are added.}
\label{fig:scalability}
\end{figure*}
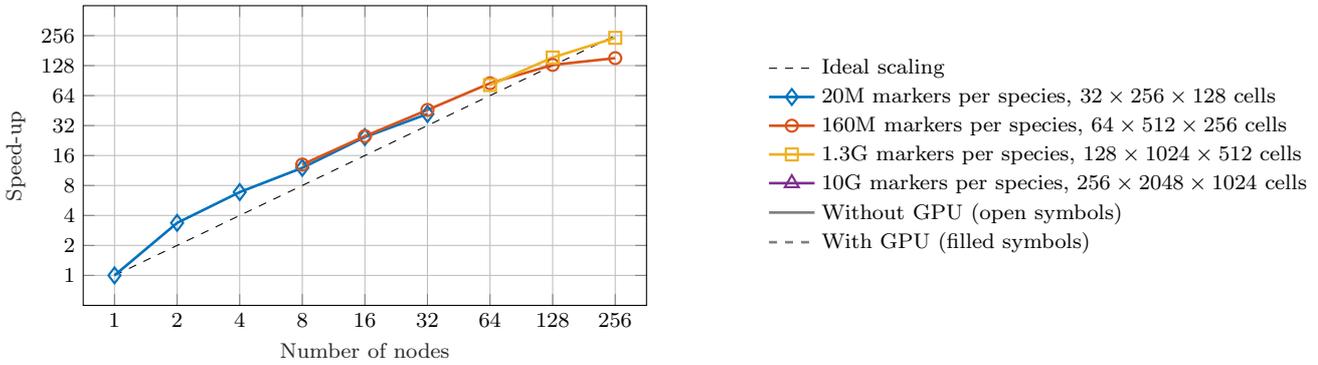
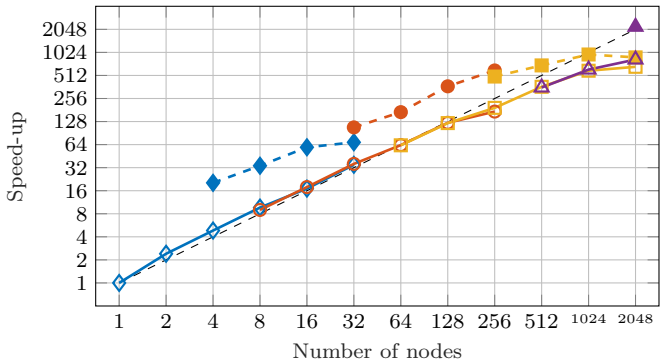
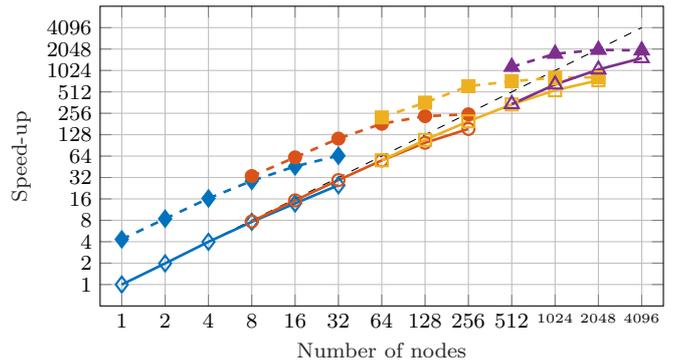

Figure \ref{fig:scalability} shows the parallel scalability of ORB5 up to 256 nodes on Marconi (\ref{fig:scalability_marconi}), up to 2048 nodes of Piz Daint with or without using the GPUs (\ref{fig:scalability_daint}) and up to  4096 compute nodes on Summit machine (\ref{fig:scalability_summit}) with or without using the GPUs. It is worth noticing that the code could run up to more than 24000 GPUs in the largest set-up on Summit.

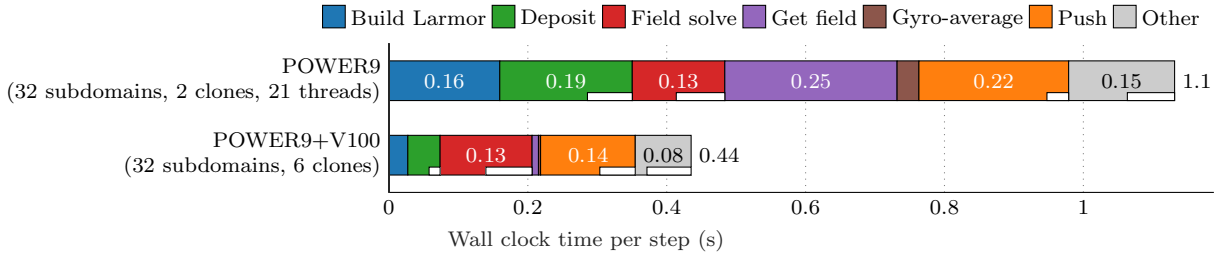
\begin{figure*}%
  \centering%
  \setlength{\figurewidth}{.9\linewidth}%
  \setlength{\figureheight}{.1196\figurewidth}%
  \footnotesize%
  \input{figures/CPU_vs_GPU_32nodes.tikz}%
  \vspace{-3mm}%
  \caption{Breakdown of ORB5 kernel timings on the small data set (20M guiding centers per species, $32\times256\times128$ cells), using 32 nodes on Summit, with or without GPUs. Timing `Other' includes in particular sources and diagnostics. The `Field solve' includes discrete Fourier transforms and backsolve operations. White bars represent time spent in MPI routines.}%
  \label{fig:CPU_vs_GPU_32nodes}%
\end{figure*}

To obtain the results of Figure \ref{fig:scalability}, four data sets are considered, which are scaled by a factor of 8 from one another for both particle and grid numbers. For each data set a strong scaling is performed, represented with a different color.
On Marconi, Figure \ref{fig:scalability_marconi}, strong and weak scalings are close to ideal up to the maximum tested (we could not run beyond 256 nodes, i.e. 12288 MPI tasks, due to limited availability of resources). The scaling appears even super-ideal, but that is an artifact due to poor performance of the single-node case, which suffers from poor cache usage due to the large memory footprint. This effect is also seen, although much smaller, on the single-node Piz Daint results, but is absent from the single-node Summit results.
Note that the single-node data set considered in section \ref{sec:node} has a much smaller grid (by a factor of 32), which was chosen to correspond to a toroidal slice of the full data set at 32 nodes, and does not suffer from this poor performance problem on any of the architectures considered.
Increasing the number of nodes for CPU-only cases on Piz Daint, Figure \ref{fig:scalability_daint} and Summit, Figure \ref{fig:scalability_summit}, the excellent strong scaling properties are confirmed for all data sets up to 24576 CPU cores on Piz Daint, respectively 172032 CPU cores on Summit.

For all data sets and number of nodes considered on both Piz Daint and Summit platforms, the use of GPU is significantly accelerating the code.
The smallest data set (20M markers per species, $32\times256\times128$ cells,  blue symbols and lines on Figure \ref{fig:scalability_summit}) on a single Summit node runs 4.3 times faster using its 6 GPUs than with its CPUs only. This factor is slightly different than the one found in section \ref{sec:arch} since the grid resolution is not the same.
The strong scalability of this case without GPUs is nearly ideal up to 8 nodes, and then slightly decreases to reach a parallel efficiency of 78\% at 32 nodes with respect to 1 node. The parallel efficiency when using the GPUs is only 47\%. This drop in efficiency can be explained with the higher MPI communication to computing ratio when GPUs are used, combined to the impact of the non-accelerated part of the code, essentially the field solve kernel, taking a higher fraction of the wall clock time, as shown in Figure \ref{fig:CPU_vs_GPU_32nodes}. The presented breakdown of the timings refers to the small data set, but it is representative also of larger configurations, for which MPI communication gets increasingly heavier, resulting in a poorer parallel scalability compared to the CPU only runs.
For instance, the second largest case (1.3G markers per species, $128\times1024\times512$ cells, yellow symbols and lines on Figure \ref{fig:scalability_summit}) running on 2048 nodes does not benefit much from the use of GPUs because it is dominated by communications. On the largest case (10G markers per species, $256\times2048\times1024$ cells, violet symbols and lines on Figure \ref{fig:scalability_summit}), the parallel efficiency of the CPU-only version at 4096 nodes with respect to 512 nodes is still 55\%.

Between the first points of each strong scalability test, the problem size and the used computational resources have been scaled by a factor of 8 (twice more cells in each direction and 8 times more markers), hence they can be used to evaluate weak scalability. On Summit, weak scalability is linear for the first three tests, while it drops on the largest test, both for CPU-only and using the GPUs, due to the big communication overhead when large system configurations are used. The largest case is 512 times heavier than the smallest one and it is 33\% slower on 512 nodes than the small one on 1 node with CPU only, or 48\% slower using the GPUs.

The single node performance of Piz Daint on the smallest case (Figure \ref{fig:scalability_daint}) without GPU, is about 4 times lower than on Summit, which is to be expected because it embeds a 12 cores Haswell CPU versus the $2\times21$ cores of the POWER9 on Summit. Due to memory constraints, at least 4 nodes are required for this case to fit the GPU memory. Using the GPUs on Piz Daint brings around a factor 4 speed-up on the linear phases of the strong scalings, leading to a performance similar to Summit without GPUs at the same number of nodes. Considering the number of nodes, the overall weak and strong scalability on Piz Daint appears to be even slightly better than that of Summit. However, one should consider that Piz Daint nodes are much lighter (less CPU cores and GPUs per node) than Summit ones, so the number of MPI tasks (hence the associated overhead) is smaller.

\begin{figure*}%
  \centering%
  \setlength{\figurewidth}{.6\linewidth}%
  \setlength{\figureheight}{.618\figurewidth}%
  \footnotesize%
  \input{figures/scalability_gpu.tikz}%
  \vspace{-3mm}%
  \caption{Comparison of strong scalings on Summit (cross symbols) and Piz Daint (circles symbols) supercomputers using GPUs. Data is the same as in Figure \ref{fig:scalability} with the total number of GPUs along the x-axis. Speed-up is normalized to the smallest case on 4 GPUs of Piz Daint, which takes 5.7s per time step.}%
  \label{fig:scalability_gpu}%
\end{figure*}
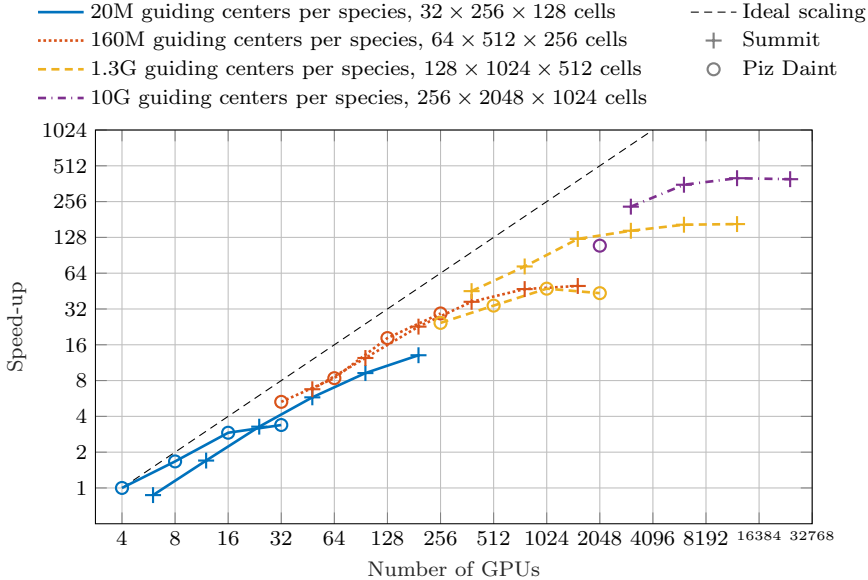

This is further shown in Figure \ref{fig:scalability_gpu}, by comparing Summit and Piz Daint scalings with respect to the number of MPI tasks instead of the number of nodes, for the GPU cases only (note that there is one MPI task per GPU). The fact that 4 Daint P100 GPUs perform better than 6 Summit V100 GPUs in the smallest case is due to the usage of 4 subdomains instead of 6 clones. However, when the computing resources grow, Piz Daint strong scalability tends to degrade earlier than that of Summit. This is because Summit benefits from more intra-node MPI communication while Piz Daint uses inter-node MPI communication. This is particularly clear comparing the scalability of the two systems for the second largest test (yellow markers).

\section{Discussion and Conclusions}
\label{sec:conclusion}

The tests presented in Section \ref{sec:results} allowed us to investigate the performance of the ORB5 code on different state-of-the-art HPC architectures, in typical production simulation setups, without any architecture-specific custom optimization (except those provided by the compiler), hence in a typical user-based scenario, using the same source code.

The parallel performance can be optimized at the application level, thanks to the different options supported by ORB5 for parallelism (domain decomposition and cloning and multithreading), whose tuning allows one to identify the most effective combination of computing resources. Often this is not driven only by the performance, but also by the memory usage and data locality. For instance, the six GPUs on a Summit fat node can be effectively exploited by dividing the domain into two subdomains (one per socket), replicated into three clones. Since the domain decomposition requires a number of MPI tasks power of two, without clones it would have been impossible to exploit all the six available GPUs. At the same time, cloning allows one to reduce the data size to fit the GPU memory.

As a general outcome, we have found that single-node parallelization optimization shows that subdomain decomposition should be prioritized, followed by OpenMP multithreading, and finally domain cloning. MPI parallelization is also useful to guarantee memory locality on multi-socket machines, avoiding to have OpenMP threads expanding over more than one socket. This priority hierarchy may be different for other simulation parameters and higher number of nodes, but it anyway provides insight for a first guess before fine-tuning optimization. The usage of OpenMP multithreading guarantees acceptable performance up to a few tens of threads both on the Skylake and on the POWER9 architectures, provided affinity is enforced and the socket boundaries are not crossed.

As expected, on a single node, the usage of the GPU leads to an important speed-up compared to non-accelerated nodes. However, such speed-up is far from being the nominal one. This is due to two main concurrent factors: the presence of the non-accelerated field solver and the overhead due to CPU-GPU data movement. The former is planned to be solved in the coming releases of the code. Since originally the solver represented a negligible part of the computation, it was left aside while focusing on the multithreaded implementation of the most time-consuming components. With all other code components accelerated, the field solver becomes an actual bottleneck and now needs to be ported on the GPU. Once this will be accomplished, also the data-movement problem will be strongly alleviated, since charge densities and forces will be permanently resident on the GPU, with no more need of transferring them at each time iteration. Further data-movement overheads due, for example, to the calculation of diagnostics or to MPI communication are expected to be effectively hidden through asynchronous processes, or through the adoption of GPUDirect.

Numerical kernels could be further optimized on the GPU by rewriting them in Cuda. To ensure the code remains future-proof, only the parts exempt from physics should be considered, such as particle deposition and field evaluation, particle sorting and field solver (all the physics is in the matrix assembly). We know from the PIC-engine studies \cite{Hariri2016} that such work could bring a few tens percents speed-up, but it is not our priority as it would affect kernels that are either taking a small fraction of the wall clock time, or planned to be upgraded in Fortran.

The challenge of accelerating PIC's critical operations of particle-to-field and field-to-particle has also been tackled introducing new algorithms using particle sorting. Even though the overall gain is not yet positive, the `deposition' and `get field' operations are significantly faster on CPU and GPU architectures. Previous studies done with the PIC-engine \cite{Hariri2016,Jocksch2015} had shown better performance with the sorting because they did not benefit from the hardware \texttt{atomic} implementation brought by the newest generations of GPUs in their reference simulations without sorting. In the future, if the cost of sorting could be reduced, a net performance gain would be achieved on ORB5. One could for instance track markers leaving cells in order to avoid full sort at each iteration.

Scalability has been investigated focusing on the two GPU-equipped systems, for which it represents a major challenge. When CPU-only tests are run, both Summit and Piz Daint show remarkable linear strong scalability up to very large configurations, losing efficiency only at the largest set-up, as shown also looking at weak scaling. When GPUs are used, the ratio between computing and communication time drops, and good strong scalability is much harder to achieve. Our tests show that ORB5 can take advantage of the Summit fat nodes architecture, strong scaling efficiently to large numbers of GPUs, with improved scalability compared to Piz Daint, that has a larger MPI communication overhead, each node having only one GPU. A second advantage of Summit is that bigger problem sizes can be handled by a single node, with an overall improvement of weak scaling. On the other hand, the Summit node resulted to be less flexible than the ``fine grained'' Piz Daint node. In order to fully exploit the resources of the Summit node we had to make use of clones, which penalized the performance compared to Piz Daint, where pure domain decomposition could be used. This resulted in better performance of Piz Daint when few GPUs (nodes) are used.

Overall, ORB5 has proved to be a prime example of a code designed for the coming generations of heterogeneous HPC architectures. Based on MPI, OpenMP and OpenACC the same codebase can be used efficiently on distributed systems as well as on multi/many threads devices. It can  exploit state-of-the-art supercomputing systems up to the largest scales, and could be tested up to more than 24000 GPUs. It is easily portable, supporting the Intel, PGI, Cray and GNU (not shown in the paper) compilers, strongly facilitating the work of the scientific user, and making accessible a broad spectrum of HPC solutions to effectively perform large scale simulations tackling the most challenging problems in plasma physics.
The directive-based tools used in this work can be considered as mature enough and widely available for other legacy codes needing similar refactoring. Among all features needed for this project, only one was problematic, namely OpenACC's GPUDirect. This way of exchanging data directly from a GPU to another was not yet fully supported by the network constructor, but it is expected to be within the next couple of years. Moreover, the only missing feature we had wished to have when porting the code to GPUs, namely OpenACC support for reduction over arrays, is now provided by standard 2.7.

\section*{Acknowledgments}
Simulations presented in this paper were performed at the Oak Ridge Leadership Computing Facility, which is a DOE Office of Science User Facility supported under Contract DE-AC05-00OR22725, Project ID is PHY137, at CSCS (Swiss National Supercomputing Center), in the User Lab program, project ID s909, and at CINECA (Italian National HPC facility), in the EUROfusion program, project ID FUA23. We acknowledge all those facilities.

We are also grateful to Stephane Ethier from PPPL (Princeton Plasma Physics Laboratory) for very fruitful discussions about porting production codes to GPUs.

This work has been carried out within the framework of the EUROfusion Consortium and has received funding from the Euratom research and training program 2014-2018 and 2019-2020 under grant agreement No 633053. The views and opinions expressed herein do not necessarily reflect those of the European Commission.
This work has been partly supported by the Swiss National Science Foundation.

\section*{Bibliography}
\bibliographystyle{unsrt} 
\bibliography{ORB5}

\end{document}

%% file: figures/MPI_vs_OpenMP.tikz
%
\definecolor{mycolor1}{rgb}{0.92900,0.69400,0.12500}%
\definecolor{mycolor2}{rgb}{0.85000,0.32500,0.09800}%
\definecolor{mycolor3}{rgb}{0.00000,0.44700,0.74100}%
\begin{tikzpicture}
\begin{axis}[%
bar width=0.7,
xmin=-23,
xmax=51.66144594625,
xtick={ 0, 10, 20, 30, 40, 50},
xlabel style={font=\color{white!15!black}},
xlabel={Wall clock time per step (s)},
ymin=0.5,
ymax=7.5,
ytick={1,2,3,4,5,6,7},
yticklabels={\empty},
axis background/.style={fill=white},
axis line style={draw=none},
axis x line*=bottom,
axis y line*=left,
legend style={at={(1,1.03)}, anchor=south east, legend columns=7, legend cell align=left, align=left, draw=white!15!black, draw=none},
width=\figurewidth,
height=\figureheight,
scale only axis,
area legend
]
\draw[draw=none, fill=mycolor1!30!white] (axis cs:-23,0.5) rectangle (axis cs:51.66144594625,2.5);
\node[fill=white, right, align=left, draw=mycolor1]
at (axis cs:-22.5,1.521) {4 subdomains};
\draw[draw=none, fill=mycolor2!30!white] (axis cs:-23,2.5) rectangle (axis cs:51.66144594625,4.5);
\node[fill=white, right, align=left, draw=mycolor2]
at (axis cs:-22.5,3.521) {2 subdomains};
\draw[draw=none, fill=mycolor3!30!white] (axis cs:-23,4.5) rectangle (axis cs:51.66144594625,7.5);
\node[fill=white, right, align=left, draw=mycolor3]
at (axis cs:-22.5,6.021) {1 subdomain};
\addplot [color=gray, dotted, forget plot]
  table[row sep=crcr]{%
10	0.5\\
10	7.5\\
};
\addplot [color=gray, dotted, forget plot]
  table[row sep=crcr]{%
20	0.5\\
20	7.5\\
};
\addplot [color=gray, dotted, forget plot]
  table[row sep=crcr]{%
30	0.5\\
30	7.5\\
};
\addplot [color=gray, dotted, forget plot]
  table[row sep=crcr]{%
40	0.5\\
40	7.5\\
};
\addplot [color=gray, dotted, forget plot]
  table[row sep=crcr]{%
50	0.5\\
50	7.5\\
};
\addplot [color=black, forget plot]
  table[row sep=crcr]{%
0	0.5\\
51.66144594625	0.5\\
};
\addplot [color=black, forget plot]
  table[row sep=crcr]{%
0	0.5\\
0	7.5\\
};
\addplot[xbar stacked, fill=mycolor_build_larmor, draw=black, my xbar legend] table[row sep=crcr] {%
2.39460666666667	1\\
2.42337958333333	2\\
2.40182625	3\\
2.50987833333333	4\\
2.85165208333333	5\\
2.40307416666667	6\\
2.46267125	7\\
};
\addplot[forget plot, color=white!15!black] table[row sep=crcr] {%
0	0.5\\
0	7.5\\
};
\addlegendentry{Build Larmor}

\addplot[xbar stacked, fill=mycolor_depos, draw=black, my xbar legend] table[row sep=crcr] {%
5.13234125	1\\
7.48068333333333	2\\
5.18281666666667	3\\
8.2058625	4\\
5.54647916666667	5\\
5.727875	6\\
9.549925	7\\
};
\addplot[forget plot, color=white!15!black] table[row sep=crcr] {%
0	0.5\\
0	7.5\\
};
\addlegendentry{Deposit}

\addplot[xbar stacked, fill=mycolor_solver, draw=black, my xbar legend] table[row sep=crcr] {%
1.77012958333333	1\\
5.4873625	2\\
2.83238333333333	3\\
9.07142916666667	4\\
5.52534166666667	5\\
5.40854166666667	6\\
16.848075	7\\
};
\addplot[forget plot, color=white!15!black] table[row sep=crcr] {%
0	0.5\\
0	7.5\\
};
\addlegendentry{Field solve}

\addplot[xbar stacked, fill=mycolor_getfield, draw=black, my xbar legend] table[row sep=crcr] {%
8.3039	1\\
12.0287541666667	2\\
7.75599583333333	3\\
12.3941333333333	4\\
8.95720416666667	5\\
7.6748125	6\\
12.1793666666667	7\\
};
\addplot[forget plot, color=white!15!black] table[row sep=crcr] {%
0	0.5\\
0	7.5\\
};
\addlegendentry{Get field}

\addplot[xbar stacked, fill=mycolor_gyroavg, draw=black, my xbar legend] table[row sep=crcr] {%
0.705229166666667	1\\
0.8539625	2\\
0.6749	3\\
0.8579625	4\\
1.14840833333333	5\\
0.696858333333333	6\\
0.844091666666667	7\\
};
\addplot[forget plot, color=white!15!black] table[row sep=crcr] {%
0	0.5\\
0	7.5\\
};
\addlegendentry{Gyro-average}

\addplot[xbar stacked, fill=mycolor_push, draw=black, my xbar legend] table[row sep=crcr] {%
4.516897541666670  	1\\
3.199216333333333		2\\
4.380810833333330		3\\
3.196931916666663		4\\
4.89769333333333	5\\
3.77112125	6\\
3.16651	7\\
};
\addplot[forget plot, color=white!15!black] table[row sep=crcr] {%
0	0.5\\
0	7.5\\
};
\addlegendentry{Push}

\addplot[xbar stacked, fill=white!80!black, draw=black, my xbar legend] table[row sep=crcr] {%
 1.938661275000003 1\\
 2.034870474999997 2\\
 1.853127949999997 3\\
 2.945238287500003 4\\
 2.449100533333330 5\\
 2.167627558333330 6\\
 4.150737508333340 7\\
};
\addplot[forget plot, color=white!15!black] table[row sep=crcr] {%
0	0.5\\
0	7.5\\
};
\addlegendentry{Other}

\node[left, align=right]
at (axis cs:0,1) {1 clone, 12 threads};
\node[left, align=right]
at (axis cs:0,2) {12 clones, 1 thread};
\node[left, align=right]
at (axis cs:0,3) {1 clone, 24 threads};
\node[left, align=right]
at (axis cs:0,4) {24 clones, 1 thread};
\node[left, align=right]
at (axis cs:0,5) {1 clone, 48 threads};
\node[left, align=right]
at (axis cs:0,6) {2 clones, 24 threads};
\node[left, align=right]
at (axis cs:0,7) {48 clones, 1 thread};
\node[align=center, font=\color{white}]
at (axis cs:1.197,1) {2.4};
\node[align=center, font=\color{white}]
at (axis cs:1.212,2) {2.4};
\node[align=center, font=\color{white}]
at (axis cs:1.201,3) {2.4};
\node[align=center, font=\color{white}]
at (axis cs:1.255,4) {2.5};
\node[align=center, font=\color{white}]
at (axis cs:1.426,5) {2.9};
\node[align=center, font=\color{white}]
at (axis cs:1.202,6) {2.4};
\node[align=center, font=\color{white}]
at (axis cs:1.231,7) {2.5};
\draw[fill=white, draw=black] (axis cs:7.518302875,0.65) rectangle (axis cs:7.52694791666667,0.79);
\node[align=center, font=\color{white}]
at (axis cs:4.961,1) {5.1};
\draw[fill=white, draw=black] (axis cs:8.59385333333333,1.65) rectangle (axis cs:9.90406291666667,1.79);
\node[align=center, font=\color{white}]
at (axis cs:6.164,2) {7.5};
\draw[fill=white, draw=black] (axis cs:7.58084807083333,2.65) rectangle (axis cs:7.58464291666667,2.79);
\node[align=center, font=\color{white}]
at (axis cs:4.993,3) {5.2};
\draw[fill=white, draw=black] (axis cs:9.03415125,3.65) rectangle (axis cs:10.7157408333333,3.79);
\node[align=center, font=\color{white}]
at (axis cs:6.613,4) {8.2};
\draw[fill=white, draw=black] (axis cs:8.39742032083333,4.65) rectangle (axis cs:8.39813125,4.79);
\node[align=center, font=\color{white}]
at (axis cs:5.625,5) {5.5};
\draw[fill=white, draw=black] (axis cs:8.11348145833333,5.65) rectangle (axis cs:8.13094916666667,5.79);
\node[align=center, font=\color{white}]
at (axis cs:5.267,6) {5.7};
\draw[fill=white, draw=black] (axis cs:9.42478625,6.65) rectangle (axis cs:12.01259625,6.79);
\node[align=center, font=\color{white}]
at (axis cs:7.238,7) {9.5};
\draw[fill=white, draw=black] (axis cs:9.266588625,0.65) rectangle (axis cs:9.2970775,0.79);
\node[align=center, font=\color{white}]
at (axis cs:8.412,1) {1.8};
\draw[fill=white, draw=black] (axis cs:14.60571625,1.65) rectangle (axis cs:15.3914254166667,1.79);
\node[align=center, font=\color{white}]
at (axis cs:12.648,2) {5.5};
\draw[fill=white, draw=black] (axis cs:10.3972098083333,2.65) rectangle (axis cs:10.41702625,2.79);
\node[align=center, font=\color{white}]
at (axis cs:9.001,3) {2.8};
\draw[fill=white, draw=black] (axis cs:18.61026375,3.65) rectangle (axis cs:19.78717,3.79);
\node[align=center, font=\color{white}]
at (axis cs:15.251,4) {9.1};
\draw[fill=white, draw=black] (axis cs:13.9081636041667,4.65) rectangle (axis cs:13.9234729166667,4.79);
\node[align=center, font=\color{white}]
at (axis cs:11.161,5) {5.5};
\draw[fill=white, draw=black] (axis cs:13.5085470833333,5.65) rectangle (axis cs:13.5394908333333,5.79);
\node[align=center, font=\color{white}]
at (axis cs:10.835,6) {5.4};
\draw[fill=white, draw=black] (axis cs:26.8511745833333,6.65) rectangle (axis cs:28.86067125,6.79);
\node[align=center, font=\color{white}]
at (axis cs:20.437,7) {17};
\node[align=center, font=\color{white}]
at (axis cs:13.449,1) {8.3};
\node[align=center, font=\color{white}]
at (axis cs:21.406,2) {12};
\node[align=center, font=\color{white}]
at (axis cs:14.295,3) {7.8};
\node[align=center, font=\color{white}]
at (axis cs:25.984,4) {12};
\node[align=center, font=\color{white}]
at (axis cs:18.402,5) {9};
\node[align=center, font=\color{white}]
at (axis cs:17.377,6) {7.7};
\node[align=center, font=\color{white}]
at (axis cs:34.95,7) {12};
\node[align=center, font=\color{white}]
at (axis cs:20.505,1) {4.5};
\node[align=center, font=\color{white}]
at (axis cs:29.834,2) {3.2};
\node[align=center, font=\color{white}]
at (axis cs:21.038,3) {4.4};
\node[align=center, font=\color{white}]
at (axis cs:34.656,4) {3.2};
\node[align=center, font=\color{white}]
at (axis cs:26.478,5) {4.9};
\node[align=center, font=\color{white}]
at (axis cs:23.797,6) {3.8};
\node[align=center, font=\color{white}]
at (axis cs:43.467,7) {3.2};
\draw[fill=white, draw=black] (axis cs:22.8119406666667,0.65) rectangle (axis cs:22.8231042083333,0.79);
\draw[fill=white, draw=black] (axis cs:31.3840775833333,1.65) rectangle (axis cs:31.4733584166667,1.79);
\draw[fill=white, draw=black] (axis cs:23.2222856666667,2.65) rectangle (axis cs:23.2287329166667,2.79);
\draw[fill=white, draw=black] (axis cs:36.1896256666667,3.65) rectangle (axis cs:36.23619775,3.79);
\draw[fill=white, draw=black] (axis cs:24.75192315,0.65) rectangle (axis cs:24.7617654833333,0.79);
\draw[fill=white, draw=black] (axis cs:33.2500118458333,1.65) rectangle (axis cs:33.5082288916667,1.79);
\draw[fill=white, draw=black] (axis cs:25.0801437125,2.65) rectangle (axis cs:25.0818608666667,2.79);
\draw[fill=white, draw=black] (axis cs:38.41607315,3.65) rectangle (axis cs:39.1814360375,3.79);
\draw[fill=white, draw=black] (axis cs:31.3758588166667,4.65) rectangle (axis cs:31.3758792833333,4.79);
\draw[fill=white, draw=black] (axis cs:27.8473271291667,5.65) rectangle (axis cs:27.849910475,5.79);
\draw[fill=white, draw=black] (axis cs:48.0872429375,6.65) rectangle (axis cs:49.2013770916667,6.79);
\node[right, align=left]
at (axis cs:49.201,7) { 49};
\node[right, align=left]
at (axis cs:27.85,6) { 28 (57\%)};
\node[right, align=left]
at (axis cs:31.376,5) { 31 (64\%)};
\node[right, align=left]
at (axis cs:39.181,4) { 39 (80\%)};
\node[right, align=left]
at (axis cs:25.082,3) { 25 (51\%)};
\node[right, align=left]
at (axis cs:33.508,2) { 34 (68\%)};
\node[right, align=left]
at (axis cs:24.762,1) { 25 (50\%)};
\end{axis}
\end{tikzpicture}%

%% file: figures/architecture_comparison.tikz
%
\begin{tikzpicture}

\begin{axis}[%
bar width=0.533,
xmin=-11.2073,
xmax=34.60245516,
xtick={ 0,  5, 10, 15, 20, 25, 30},
xlabel style={font=\color{white!15!black}},
xlabel={Wall clock time per step (s)},
ymin=0.5,
ymax=4.5,
ytick={0.5,1,1.5,2,2.5,3,3.5,4,4.5},
yticklabels={\empty},
axis background/.style={fill=white},
axis x line*=bottom,
axis y line*=left,
legend style={at={(1,1.03)}, anchor=south east, legend columns=7, legend cell align=left, align=left, draw=white!15!black, draw=none},
width=\figurewidth,
height=\figureheight,
axis line style={draw=none},
tick style={draw=none},
scale only axis
]
\addplot [color=gray, dotted, forget plot]
  table[row sep=crcr]{%
5	0.5\\
5	4.5\\
};
\addplot [color=gray, dotted, forget plot]
  table[row sep=crcr]{%
10	0.5\\
10	4.5\\
};
\addplot [color=gray, dotted, forget plot]
  table[row sep=crcr]{%
15	0.5\\
15	4.5\\
};
\addplot [color=gray, dotted, forget plot]
  table[row sep=crcr]{%
20	0.5\\
20	4.5\\
};
\addplot [color=gray, dotted, forget plot]
  table[row sep=crcr]{%
25	0.5\\
25	4.5\\
};
\addplot [color=gray, dotted, forget plot]
  table[row sep=crcr]{%
30	0.5\\
30	4.5\\
};
\addplot [color=black, forget plot]
  table[row sep=crcr]{%
0	0.5\\
34.60245516	0.5\\
};
\addplot [color=black, forget plot]
  table[row sep=crcr]{%
0	0.5\\
0	4.5\\
};
\addplot[xbar stacked, fill=mycolor_build_larmor, draw=black, my xbar legend] table[row sep=crcr] {%
0.904865	1\\
0.732745	2\\
4.820674	3\\
2.39460666666667	4\\
};
\addplot[forget plot, color=white!15!black] table[row sep=crcr] {%
0	0.5\\
0	4.5\\
};
\addlegendentry{Build Larmor}

\addplot[xbar stacked, fill=mycolor_depos, draw=black, my xbar legend] table[row sep=crcr] {%
2.23038	1\\
1.39350666666667	2\\
7.08454	3\\
5.13234125	4\\
};
\addplot[forget plot, color=white!15!black] table[row sep=crcr] {%
0	0.5\\
0	4.5\\
};
\addlegendentry{Deposit}

\addplot[xbar stacked, fill=mycolor_solver, draw=black, my xbar legend] table[row sep=crcr] {%
2.99827666666667	1\\
2.92835666666667	2\\
2.34156	3\\
1.77012958333333	4\\
};
\addplot[forget plot, color=white!15!black] table[row sep=crcr] {%
0	0.5\\
0	4.5\\
};
\addlegendentry{Field solve}

\addplot[xbar stacked, fill=mycolor_getfield, draw=black, my xbar legend] table[row sep=crcr] {%
0.687096666666667	1\\
0.357236	2\\
9.319804	3\\
8.3039	4\\
};
\addplot[forget plot, color=white!15!black] table[row sep=crcr] {%
0	0.5\\
0	4.5\\
};
\addlegendentry{Get field}

\addplot[xbar stacked, fill=mycolor_gyroavg, draw=black, my xbar legend] table[row sep=crcr] {%
0.149981	1\\
0.0714523333333333	2\\
0.748409	3\\
0.705229166666667	4\\
};
\addplot[forget plot, color=white!15!black] table[row sep=crcr] {%
0	0.5\\
0	4.5\\
};
\addlegendentry{Gyro-average}

\addplot[xbar stacked, fill=mycolor_push, draw=black, my xbar legend] table[row sep=crcr] {%
0.834075533333333    1\\
0.484826533333333    2\\
5.719998900000000    3\\
4.516897541666670    4\\
};
\addplot[forget plot, color=white!15!black] table[row sep=crcr] {%
0	0.5\\
0	4.5\\
};
\addlegendentry{Push}

\addplot[xbar stacked, fill=white!80!black, draw=black, my xbar legend] table[row sep=crcr] {%
1.032616879170000	1\\
0.719313566666667	2\\
2.919733300000000	3\\
1.938661275000003	4\\
};
\addplot[forget plot, color=white!15!black] table[row sep=crcr] {%
0	0.5\\
0	4.5\\
};
\addlegendentry{Other}

\node[left, align=right]
at (axis cs:0,1) {Haswell+P100\\(2 subdomains, 3 clones)};
\node[left, align=right]
at (axis cs:0,2) {POWER9+V100\\(2 subdomains, 3 clones)};
\node[left, align=right]
at (axis cs:0,3) {POWER9\\(2 subdomains, 21 threads)};
\node[left, align=right]
at (axis cs:0,4) {Skylake\\(4 subdomains, 12 threads)};
\node[align=center, font=\color{white}]
at (axis cs:2.41,3) {4.8};
\node[align=center, font=\color{white}]
at (axis cs:1.197,4) {2.4};
\draw[fill=white, draw=black] (axis cs:2.656974,0.7335) rectangle (axis cs:3.135245,0.84);
\node[align=center, font=\color{white}]
at (axis cs:2.02,1) {2.2};
\draw[fill=white, draw=black] (axis cs:1.77203233333333,1.7335) rectangle (axis cs:2.12625166666667,1.84);
\node[align=center, font=\color{white}]
at (axis cs:1.429,2) {1.4};
\draw[fill=white, draw=black] (axis cs:11.6796504,2.7335) rectangle (axis cs:11.905214,2.84);
\node[align=center, font=\color{white}]
at (axis cs:8.363,3) {7.1};
\draw[fill=white, draw=black] (axis cs:7.518302875,3.7335) rectangle (axis cs:7.52694791666667,3.84);
\node[align=center, font=\color{white}]
at (axis cs:4.961,4) {5.1};
\draw[fill=white, draw=black] (axis cs:5.61013433333333,0.7335) rectangle (axis cs:6.13352166666667,0.84);
\node[align=center, font=\color{white}]
at (axis cs:4.634,1) {3};
\draw[fill=white, draw=black] (axis cs:4.58496266666667,1.7335) rectangle (axis cs:5.05460833333333,1.84);
\node[align=center, font=\color{white}]
at (axis cs:3.59,2) {2.9};
\draw[fill=white, draw=black] (axis cs:13.9380923,2.7335) rectangle (axis cs:14.246774,2.84);
\node[align=center, font=\color{white}]
at (axis cs:13.076,3) {2.3};
\draw[fill=white, draw=black] (axis cs:9.266588625,3.7335) rectangle (axis cs:9.2970775,3.84);
\node[align=center, font=\color{white}]
at (axis cs:8.412,4) {1.8};
\node[align=center, font=\color{white}]
at (axis cs:18.907,3) {9.3};
\node[align=center, font=\color{white}]
at (axis cs:13.449,4) {8.3};
\draw[fill=white, draw=black] (axis cs:7.72849333333333,0.7335) rectangle (axis cs:7.74398786666667,0.84);
\draw[fill=white, draw=black] (axis cs:5.90575303333333,1.7335) rectangle (axis cs:5.91477486666667,1.84);
\node[align=center, font=\color{white}]
at (axis cs:27.188,3) {5.7};
\node[align=center, font=\color{white}]
at (axis cs:20.305,4) {4.5};
\draw[fill=white, draw=black] (axis cs:7.77858046666667,0.7335) rectangle (axis cs:7.80467486666667,0.84);
\draw[fill=white, draw=black] (axis cs:5.95081206666667,1.7335) rectangle (axis cs:5.9681232,1.84);
\draw[fill=white, draw=black] (axis cs:29.8116839,2.7335) rectangle (axis cs:30.0349859,2.84);
\draw[fill=white, draw=black] (axis cs:22.8119406666667,3.7335) rectangle (axis cs:22.8231042083333,3.84);
\draw[fill=white, draw=black] (axis cs:8.67002914150333,0.7335) rectangle (axis cs:8.83729174583667,0.84);
\draw[fill=white, draw=black] (axis cs:6.52749853333333,1.7335) rectangle (axis cs:6.68743676666667,1.84);
\draw[fill=white, draw=black] (axis cs:32.8407307,2.7335) rectangle (axis cs:32.9547192,2.84);
\draw[fill=white, draw=black] (axis cs:24.75192315,3.7335) rectangle (axis cs:24.7617654833333,3.84);
\node[right, align=left]
at (axis cs:24.762,4) { 25};
\node[right, align=left]
at (axis cs:32.955,3) { 33};
\node[right, align=left]
at (axis cs:6.687,2) { 6.7};
\node[right, align=left]
at (axis cs:8.837,1) { 8.8};
\end{axis}
\end{tikzpicture}%

%% file: figures/sorting.tikz
%
\definecolor{mycolor1}{rgb}{0.85000,0.32500,0.09800}%
\definecolor{mycolor2}{rgb}{0.00000,0.44700,0.74100}%
\begin{tikzpicture}

\begin{axis}[%
bar width=0.7,
xmin=-9,
xmax=27.78754153875,
xtick={ 0,  5, 10, 15, 20, 25},
xlabel style={font=\color{white!15!black}},
xlabel={Wall clock time per step (s)},
ymin=0.5,
ymax=4.5,
ytick={0.5,1,1.5,2,2.5,3,3.5,4,4.5},
yticklabels={\empty},
axis background/.style={fill=white},
axis x line*=bottom,
axis y line*=left,
legend style={at={(1,1.03)}, anchor=south east, legend columns=8, legend cell align=left, align=left, draw=white!15!black, draw=none},
width=\figurewidth,
height=\figureheight,
axis line style={draw=none},
tick style={draw=none},
scale only axis
]
\draw[draw=none, fill=mycolor1!30!white] (axis cs:-9,0.5) rectangle (axis cs:27.78754153875,2.5);
\node[fill=white, right, align=left, draw=mycolor1]
at (axis cs:-8.5,1.5) {POWER9\\+V100};
\draw[draw=none, fill=mycolor2!30!white] (axis cs:-9,2.5) rectangle (axis cs:27.78754153875,4.5);
\node[fill=white, right, align=left, draw=mycolor2]
at (axis cs:-8.5,3.5) {Skylake};
\addplot [color=gray, dotted, forget plot]
  table[row sep=crcr]{%
5	0.5\\
5	4.5\\
};
\addplot [color=gray, dotted, forget plot]
  table[row sep=crcr]{%
10	0.5\\
10	4.5\\
};
\addplot [color=gray, dotted, forget plot]
  table[row sep=crcr]{%
15	0.5\\
15	4.5\\
};
\addplot [color=gray, dotted, forget plot]
  table[row sep=crcr]{%
20	0.5\\
20	4.5\\
};
\addplot [color=gray, dotted, forget plot]
  table[row sep=crcr]{%
25	0.5\\
25	4.5\\
};
\addplot [color=black, forget plot]
  table[row sep=crcr]{%
0	0.5\\
27.78754153875	0.5\\
};
\addplot [color=black, forget plot]
  table[row sep=crcr]{%
0	0.5\\
0	4.5\\
};
\addplot[xbar stacked, fill=mycolor_build_larmor, draw=black, my xbar legend] table[row sep=crcr] {%
0.733293333333333	1\\
0.732745	2\\
2.38341125	3\\
2.39460666666667	4\\
};
\addplot[forget plot, color=white!15!black] table[row sep=crcr] {%
0	0.5\\
0	4.5\\
};
\addlegendentry{Build Larmor}

\addplot[xbar stacked, fill=mycolor_sort, draw=black, my xbar legend] table[row sep=crcr] {%
0.745106333333333	1\\
0	2\\
7.27912083333333	3\\
0	4\\
};
\addplot[forget plot, color=white!15!black] table[row sep=crcr] {%
0	0.5\\
0	4.5\\
};
\addlegendentry{Sort}

\addplot[xbar stacked, fill=mycolor_depos, draw=black, my xbar legend] table[row sep=crcr] {%
0.883303333333333	1\\
1.39350666666667	2\\
2.14394875	3\\
5.13234125	4\\
};
\addplot[forget plot, color=white!15!black] table[row sep=crcr] {%
0	0.5\\
0	4.5\\
};
\addlegendentry{Deposit}

\addplot[xbar stacked, fill=mycolor_solver, draw=black, my xbar legend] table[row sep=crcr] {%
2.95183	1\\
2.92835666666667	2\\
1.75423791666667	3\\
1.77012958333333	4\\
};
\addplot[forget plot, color=white!15!black] table[row sep=crcr] {%
0	0.5\\
0	4.5\\
};
\addlegendentry{Field solve}

\addplot[xbar stacked, fill=mycolor_getfield, draw=black, my xbar legend] table[row sep=crcr] {%
0.223329333333333	1\\
0.357236	2\\
5.83405	3\\
8.3039	4\\
};
\addplot[forget plot, color=white!15!black] table[row sep=crcr] {%
0	0.5\\
0	4.5\\
};
\addlegendentry{Get field}

\addplot[xbar stacked, fill=mycolor_gyroavg, draw=black, my xbar legend] table[row sep=crcr] {%
0.071366	1\\
0.0714523333333333	2\\
0.684770833333333	3\\
0.705229166666667	4\\
};
\addplot[forget plot, color=white!15!black] table[row sep=crcr] {%
0	0.5\\
0	4.5\\
};
\addlegendentry{Gyro-average}

\addplot[xbar stacked, fill=mycolor_push, draw=black, my xbar legend] table[row sep=crcr] {%
0.517776766666667	1\\
0.484826533333333	2\\
4.43862791666667	3\\
4.51689754166667	4\\
};
\addplot[forget plot, color=white!15!black] table[row sep=crcr] {%
0	0.5\\
0	4.5\\
};
\addlegendentry{Push}

\addplot[xbar stacked, fill=white!80!black, draw=black, my xbar legend] table[row sep=crcr] {%
0.7596196	1\\
0.719313566666667	2\\
1.946157775	3\\
1.938661275	4\\
};
\addplot[forget plot, color=white!15!black] table[row sep=crcr] {%
0	0.5\\
0	4.5\\
};
\addlegendentry{Other}

\node[left, align=right]
at (axis cs:0,1) {sorting};
\node[left, align=right]
at (axis cs:0,2) {no sorting};
\node[left, align=right]
at (axis cs:0,3) {sorting};
\node[left, align=right]
at (axis cs:0,4) {no sorting};
\node[align=center, font=\color{white}]
at (axis cs:1.192,3) {2.4};
\node[align=center, font=\color{white}]
at (axis cs:1.197,4) {2.4};
\node[align=center, font=\color{white}]
at (axis cs:6.023,3) {7.3};
\draw[fill=white, draw=black] (axis cs:2.01531733333333,0.65) rectangle (axis cs:2.361703,0.79);
\draw[fill=white, draw=black] (axis cs:1.77203233333333,1.65) rectangle (axis cs:2.12625166666667,1.79);
\node[align=center, font=\color{white}]
at (axis cs:1.429,2) {1.4};
\draw[fill=white, draw=black] (axis cs:11.795626125,2.65) rectangle (axis cs:11.8064808333333,2.79);
\node[align=center, font=\color{white}]
at (axis cs:10.735,3) {2.1};
\draw[fill=white, draw=black] (axis cs:7.518302875,3.65) rectangle (axis cs:7.52694791666667,3.79);
\node[align=center, font=\color{white}]
at (axis cs:4.961,4) {5.1};
\draw[fill=white, draw=black] (axis cs:4.85846233333333,0.65) rectangle (axis cs:5.313533,0.79);
\node[align=center, font=\color{white}]
at (axis cs:3.838,1) {3};
\draw[fill=white, draw=black] (axis cs:4.58496266666667,1.65) rectangle (axis cs:5.05460833333333,1.79);
\node[align=center, font=\color{white}]
at (axis cs:3.59,2) {3};
\draw[fill=white, draw=black] (axis cs:13.5310480416667,2.65) rectangle (axis cs:13.56071875,2.79);
\node[align=center, font=\color{white}]
at (axis cs:12.684,3) {1.8};
\draw[fill=white, draw=black] (axis cs:9.266588625,3.65) rectangle (axis cs:9.2970775,3.79);
\node[align=center, font=\color{white}]
at (axis cs:8.412,4) {1.8};
\node[align=center, font=\color{white}]
at (axis cs:16.478,3) {5.8};
\node[align=center, font=\color{white}]
at (axis cs:13.449,4) {8.3};
\draw[fill=white, draw=black] (axis cs:6.08389886666667,0.65) rectangle (axis cs:6.1260051,0.79);
\draw[fill=white, draw=black] (axis cs:5.94179023333333,1.65) rectangle (axis cs:5.9681232,1.79);
\draw[fill=white, draw=black] (axis cs:24.4978160833333,2.65) rectangle (axis cs:24.5181675,2.79);
\node[align=center, font=\color{white}]
at (axis cs:22.299,3) {4.4};
\draw[fill=white, draw=black] (axis cs:22.8073807916667,3.65) rectangle (axis cs:22.8231042083333,3.79);
\node[align=center, font=\color{white}]
at (axis cs:20.565,4) {4.5};
\draw[fill=white, draw=black] (axis cs:6.6322401,0.65) rectangle (axis cs:6.8856247,0.79);
\draw[fill=white, draw=black] (axis cs:6.4706152,1.65) rectangle (axis cs:6.68743676666667,1.79);
\draw[fill=white, draw=black] (axis cs:26.4482509416667,2.65) rectangle (axis cs:26.464325275,2.79);
\draw[fill=white, draw=black] (axis cs:24.7467151083333,3.65) rectangle (axis cs:24.7617654833333,3.79);
\node[right, align=left]
at (axis cs:24.762,4) { 25};
\node[right, align=left]
at (axis cs:26.464,3) { 26};
\node[right, align=left]
at (axis cs:6.687,2) { 6.7};
\node[right, align=left]
at (axis cs:6.886,1) { 6.9};
\end{axis}
\end{tikzpicture}%

%% file: figures/scalability_marconi.tikz
%
\definecolor{mycolor1}{rgb}{0.00000,0.44700,0.74100}%
\definecolor{mycolor2}{rgb}{0.85000,0.32500,0.09800}%
\definecolor{mycolor3}{rgb}{0.92900,0.69400,0.12500}%
\definecolor{mycolor4}{rgb}{0.49400,0.18400,0.55600}%
\begin{tikzpicture}

\begin{axis}[%
at={(0,1.02\figureheight)},
unbounded coords=jump,
xmode=log,
xmin=0.707106781186547,
xmax=362,
xtick={   1,    2,    4,    8,   16,   32,   64,  128,  256},
xticklabels={},
xticklabels={   1,    2,    4,    8,   16,   32,   64,  128,  256},
xlabel style={font=\color{white!15!black}},
xlabel={Number of nodes},
ymode=log,
ymin=0.5,
ymax=512,
ytick={   1,    2,    4,    8,   16,   32,   64,  128,  256},
yticklabels={   1,    2,    4,    8,   16,   32,   64,  128,  256},
ylabel style={font=\color{white!15!black}},
ylabel={Speed-up},
axis background/.style={fill=white},
xmajorgrids,
ymajorgrids,
width=\figurewidth,
height=\figureheight,
legend style={legend cell align=left, align=left, draw=white!15!black, at={(1.2,0.5)}, anchor=west, draw=none}
]
\addplot [color=black, dashed]
  table[row sep=crcr]{%
1	1\\
256	256\\
};
\addlegendentry{Ideal scaling}

\addplot [color=mycolor1, line width=1pt, mark=diamond, mark size=2pt]
  table[row sep=crcr]{%
1	1\\
2	3.36779444679768\\
4	6.86526446264671\\
8	12.0320473370316\\
16	24.5247137590126\\
32	41.833284474504\\
};
\addlegendentry{20M markers per species, $32\times256\times128$ cells}

\addplot [color=mycolor2, line width=1pt, mark=o, mark size=1.5pt]
  table[row sep=crcr]{%
8	12.9968175431236\\
16	25.0379247243977\\
32	45.9310211158961\\
64	85.1450176153826\\
128	130.109532470695\\
256	152.118418334804\\
};
\addlegendentry{160M markers per species, $64\times512\times256$ cells}

\addplot [color=mycolor3, line width=1pt, mark=square, mark size=1.5pt]
  table[row sep=crcr]{%
64	81.6362299798253\\
128	154.443638677608\\
256	243.387106868359\\
512	nan\\
1024	nan\\
2048	nan\\
};
\addlegendentry{1.3G markers per species, $128\times1024\times512$ cells}

\addplot [color=mycolor4, line width=1pt, mark=triangle, mark size=2pt]
  table[row sep=crcr]{%
0.5	0.5\\
};
\addlegendentry{10G markers per species, $256\times2048\times1024$ cells}

\addplot [color=gray, line width=1pt]
  table[row sep=crcr]{%
0.5	0.5\\
};
\addlegendentry{Without GPU (open symbols)}

\addplot [color=gray, dashed, line width=1pt]
  table[row sep=crcr]{%
0.5	0.5\\
};
\addlegendentry{With GPU (filled symbols)}

\end{axis}
\end{tikzpicture}%

%% file: figures/scalability_daint.tikz
%
\definecolor{mycolor1}{rgb}{0.00000,0.44700,0.74100}%
\definecolor{mycolor2}{rgb}{0.85000,0.32500,0.09800}%
\definecolor{mycolor3}{rgb}{0.92900,0.69400,0.12500}%
\definecolor{mycolor4}{rgb}{0.49400,0.18400,0.55600}%
\begin{tikzpicture}

\begin{axis}[%
unbounded coords=jump,
xmode=log,
xmin=0.707106781186547,
xmax=2896.309,
xtick={   1,    2,    4,    8,   16,   32,   64,  128,  256,  512, 1024, 2048},
xticklabels={   1,    2,    4,    8,   16,   32,   64,  128,  256,  512,\tiny 1024,\tiny 2048},
xlabel style={font=\color{white!15!black}},
xlabel={Number of nodes},
ymode=log,
ymin=0.5,
ymax=4096,
ytick={   1,    2,    4,    8,   16,   32,   64,  128,  256,  512, 1024, 2048},
yticklabels={   1,    2,    4,    8,   16,   32,   64,  128,  256,  512, 1024, 2048},
ylabel style={font=\color{white!15!black}},
ylabel={Speed-up},
axis background/.style={fill=white},
xmajorgrids,
ymajorgrids,
legend style={legend cell align=left, align=left, draw=white!15!black, at={(0.01,0.99)}, anchor=north west},
width=\figurewidth,
height=\figureheight,
]
\addplot [color=black, dashed]
  table[row sep=crcr]{%
1	1\\
2048	2048\\
};
\addlegendentry{Ideal scaling}

\addplot [color=mycolor1, dashed, line width=1pt, mark=diamond*, mark size=2pt, mark options={solid}, forget plot]
  table[row sep=crcr]{%
1	nan\\
2	nan\\
4	20.270690425773\\
8	33.8883368530397\\
16	59.055121531603\\
32	68.5861653051424\\
};
\addplot [color=mycolor2, dashed, line width=1pt, mark=*, mark size=1.5pt, mark options={solid}, forget plot]
  table[row sep=crcr]{%
8	nan\\
16	nan\\
32	107.413462044108\\
64	169.909991982702\\
128	369.340049679598\\
256	594.918918579909\\
};
\addplot [color=mycolor3, dashed, line width=1pt, mark=square*, mark size=1.5pt, mark options={solid}, forget plot]
  table[row sep=crcr]{%
64	nan\\
128	nan\\
256	495.413440888446\\
512	692.400240750461\\
1024	961.334136895846\\
2048	881.280377875902\\
};
\addplot [color=mycolor4, dashed, line width=1pt, mark=triangle*, mark size=2pt, mark options={solid}, forget plot]
  table[row sep=crcr]{%
512	nan\\
1024	nan\\
2048	2210.023843097\\
4096	nan\\
8192	nan\\
16384	nan\\
};

\addplot [color=mycolor1, line width=1pt, mark=diamond, mark size=2pt]
  table[row sep=crcr]{%
1	1\\
2	2.4036305006266\\
4	4.83156508058839\\
8	9.63503843927464\\
16	17.0510971798892\\
32	34.8847939155661\\
};
\addlegendentry{20M particles, $32\times256\times128$ cells}

\addplot [color=mycolor2, line width=1pt, mark=o, mark size=1.5pt]
  table[row sep=crcr]{%
8	8.976381747634\\
16	17.8581672978357\\
32	35.8379960746948\\
64	63.0666218086823\\
128	123.221645284419\\
256	172.818868754477\\
};
\addlegendentry{160M particles, $64\times512\times256$ cells}

\addplot [color=mycolor3, line width=1pt, mark=square, mark size=1.5pt]
  table[row sep=crcr]{%
64	62.688125317046\\
128	122.226978015448\\
256	193.246986094212\\
512	363.884329051073\\
1024	590.227747741646\\
2048	663.620608016638\\
};
\addlegendentry{1.3G particles, $128\times1024\times512$ cells}

\addplot [color=mycolor4, line width=1pt, mark=triangle, mark size=2pt]
  table[row sep=crcr]{%
512	356.040122406491\\
1024	610.929910519203\\
2048	826.3435998522\\
4096	nan\\
8192	nan\\
16384	nan\\
};
\addlegendentry{10G particles, $256\times2048\times1024$ cells}

\addplot [color=gray, line width=2.0pt, mark=+, mark options={solid, gray}]
  table[row sep=crcr]{%
0.5	0.5\\
};
\addlegendentry{CPU}

\addplot [color=gray, dashed, line width=2.0pt, mark=o, mark options={solid, gray}]
  table[row sep=crcr]{%
0.5	0.5\\
};
\addlegendentry{GPU}

\legend{}; 

\end{axis}
\end{tikzpicture}%

%% file: figures/scalability_summit.tikz
%
\definecolor{mycolor1}{rgb}{0.00000,0.44700,0.74100}%
\definecolor{mycolor2}{rgb}{0.85000,0.32500,0.09800}%
\definecolor{mycolor3}{rgb}{0.92900,0.69400,0.12500}%
\definecolor{mycolor4}{rgb}{0.49400,0.18400,0.55600}%
\begin{tikzpicture}

\begin{axis}[%
unbounded coords=jump,
xmode=log,
xmin=0.707106781186547,
xmax=5792.618,
xtick={   1,    2,    4,    8,   16,   32,   64,  128,  256,  512, 1024, 2048, 4096},
xticklabels={   1,    2,    4,    8,   16,   32,   64,  128,  256,  512, \tiny 1024,\tiny 2048,\tiny 4096},
xlabel style={font=\color{white!15!black}},
xlabel={Number of nodes},
ymode=log,
ymin=0.5,
ymax=8192,
ytick={   1,    2,    4,    8,   16,   32,   64,  128,  256,  512, 1024, 2048, 4096},
yticklabels={   1,    2,    4,    8,   16,   32,   64,  128,  256,  512, 1024, 2048,4096},
ylabel style={font=\color{white!15!black}},
ylabel={Speed-up},
xmajorgrids,
ymajorgrids,
legend style={legend cell align=left, align=left, draw=white!15!black, at={(0.5,1.03)}, anchor=south}, 
width=\figurewidth,
height=\figureheight
]
\addplot [color=black, dashed]
  table[row sep=crcr]{%
1	1\\
4096	4096\\
};
\addlegendentry{Ideal scaling}

\addplot [color=mycolor1, dashed, line width=1pt, mark=diamond*, mark size=2pt, mark options={solid}, forget plot]
  table[row sep=crcr]{%
1	4.31684209386481\\
2	8.42709354588613\\
4	16.2445571063637\\
8	28.6621559556612\\
16	45.8491273960706\\
32	64.7721040402092\\
};
\addplot [color=mycolor2, dashed, line width=1pt, mark=*, mark size=1.5pt, mark options={solid}, forget plot]
  table[row sep=crcr]{%
8	33.5330376929114\\
16	61.400103085673\\
32	112.736015310512\\
64	182.348727522264\\
128	233.212763362503\\
256	247.66979789851\\
};
\addplot [color=mycolor3, dashed, line width=1pt, mark=square*, mark size=1.5pt, mark options={solid}, forget plot]
  table[row sep=crcr]{%
64	224.17424696778\\
128	361.464785129741\\
256	615.832544931573\\
512	721.600357940723\\
1024	811.013147914338\\
2048	821.183111501069\\
};
\addplot [color=mycolor4, dashed, line width=1pt, mark=triangle*, mark size=2pt, mark options={solid}, forget plot]
  table[row sep=crcr]{%
512	1150.61717355085\\
1024	1759.07570470133\\
2048	1996.29835057099\\
4096	1959.10979721994\\
8192	nan\\
16384	nan\\
};

\addplot [color=mycolor1, line width=1pt, mark=diamond, mark size=2pt]
  table[row sep=crcr]{%
1	1\\
2	1.97968204308976\\
4	3.9831293694271\\
8	7.60955096931499\\
16	13.8172004582399\\
32	24.9121359500356\\
};
\addlegendentry{20M markers per species, $32\times256\times128$ cells}

\addplot [color=mycolor2, line width=1pt, mark=o, mark size=1.5pt]
  table[row sep=crcr]{%
8	7.71809256768448\\
16	15.2508219392296\\
32	29.4231775684498\\
64	55.6867042097775\\
128	98.5474232269694\\
256	154.214767121206\\
};
\addlegendentry{160M markers per species, $64\times512\times256$ cells}

\addplot [color=mycolor3, line width=1pt, mark=square, mark size=1.5pt]
  table[row sep=crcr]{%
64	55.8264987945034\\
128	108.176499399466\\
256	197.328218685014\\
512	341.101492796892\\
1024	538.81060480474\\
2048	745.374634241635\\
};
\addlegendentry{1.3G markers per species, $128\times1024\times512$ cells}

\addplot [color=mycolor4, line width=1pt, mark=triangle, mark size=2pt]
  table[row sep=crcr]{%
512	344.215285471903\\
1024	650.623964198573\\
2048	1056.99964013268\\
4096	1525.99736624479\\
8192	nan\\
16384	nan\\
};
\addlegendentry{10G markers per species, $256\times2048\times1024$ cells}

\addplot [color=gray, line width=1pt]
  table[row sep=crcr]{%
0.5	0.5\\
};
\addlegendentry{Without GPU (open symbols)}

\addplot [color=gray, dashed, line width=1pt]
  table[row sep=crcr]{%
0.5	0.5\\
};
\addlegendentry{With GPU (filled symbols)}

\legend{}; 

\end{axis}
\end{tikzpicture}%

%% file: figures/CPU_vs_GPU_32nodes.tikz
%
\begin{tikzpicture}

\begin{axis}[%
bar width=0.533,
xmin=-0.62,
xmax=1.18795130671875,
xtick={0, 0.2, 0.4, 0.6, 0.8,   1},
xlabel style={font=\color{white!15!black}},
xlabel={Wall clock time per step (s)},
ymin=0.5,
ymax=2.5,
ytick={0.5,1,1.5,2,2.5},
yticklabels={\empty},
axis background/.style={fill=white},
axis x line*=bottom,
axis y line*=left,
legend style={at={(1,1.03)}, anchor=south east, legend columns=7, legend cell align=left, align=left, draw=white!15!black, draw=none},
width=\figurewidth,
height=\figureheight,
axis line style={draw=none},
tick style={draw=none},
scale only axis
]
\draw[draw=none, fill=white] (axis cs:-0.45255287875,0.5) rectangle (axis cs:1.18795130671875,2.5);
\addplot [color=gray, dotted, forget plot]
  table[row sep=crcr]{%
0.2	0.5\\
0.2	2.5\\
};
\addplot [color=gray, dotted, forget plot]
  table[row sep=crcr]{%
0.4	0.5\\
0.4	2.5\\
};
\addplot [color=gray, dotted, forget plot]
  table[row sep=crcr]{%
0.6	0.5\\
0.6	2.5\\
};
\addplot [color=gray, dotted, forget plot]
  table[row sep=crcr]{%
0.8	0.5\\
0.8	2.5\\
};
\addplot [color=gray, dotted, forget plot]
  table[row sep=crcr]{%
1	0.5\\
1	2.5\\
};
\addplot [color=black, forget plot]
  table[row sep=crcr]{%
0	0.5\\
1.18795130671875	0.5\\
};
\addplot [color=black, forget plot]
  table[row sep=crcr]{%
0	0.5\\
0	2.5\\
};
\addplot[xbar stacked, fill=mycolor_build_larmor, draw=black, , my xbar legend] table[row sep=crcr] {%
0.0271659375	1\\
0.15958384375	2\\
};
\addplot[forget plot, color=white!15!black] table[row sep=crcr] {%
0	0.5\\
0	2.5\\
};
\addlegendentry{Build Larmor}

\addplot[xbar stacked, fill=mycolor_depos, draw=black, my xbar legend] table[row sep=crcr] {%
0.0465082291666667	1\\
0.1908088125	2\\
};
\addplot[forget plot, color=white!15!black] table[row sep=crcr] {%
0	0.5\\
0	2.5\\
};
\addlegendentry{Deposit}

\addplot[xbar stacked, fill=mycolor_solver, draw=black, my xbar legend] table[row sep=crcr] {%
0.1324459375	1\\
0.133251625	2\\
};
\addplot[forget plot, color=white!15!black] table[row sep=crcr] {%
0	0.5\\
0	2.5\\
};
\addlegendentry{Field solve}

\addplot[xbar stacked, fill=mycolor_getfield, draw=black, my xbar legend] table[row sep=crcr] {%
0.00920991666666667	1\\
0.24811875	2\\
};
\addplot[forget plot, color=white!15!black] table[row sep=crcr] {%
0	0.5\\
0	2.5\\
};
\addlegendentry{Get field}

\addplot[xbar stacked, fill=mycolor_gyroavg, draw=black, my xbar legend] table[row sep=crcr] {%
0.00303141666666667	1\\
0.0314259375	2\\
};
\addplot[forget plot, color=white!15!black] table[row sep=crcr] {%
0	0.5\\
0	2.5\\
};
\addlegendentry{Gyro-average}

\addplot[xbar stacked, fill=mycolor_push, draw=black, my xbar legend] table[row sep=crcr] {%
0.136329051041667	1\\
0.215613365625	2\\
};
\addplot[forget plot, color=white!15!black] table[row sep=crcr] {%
0	0.5\\
0	2.5\\
};
\addlegendentry{Push}

\addplot[xbar stacked, fill=white!80!black, draw=black, my xbar legend] table[row sep=crcr] {%
0.0804528114583333	1\\
0.1525798625	2\\
};
\addplot[forget plot, color=white!15!black] table[row sep=crcr] {%
0	0.5\\
0	2.5\\
};
\addlegendentry{Other}

\node[left, align=right]
at (axis cs:0,1) {POWER9+V100\\(32 subdomains, 6 clones)};
\node[left, align=right]
at (axis cs:0,2) {POWER9\\(32 subdomains, 2 clones, 21 threads)};
\node[align=center, font=\color{white}]
at (axis cs:0.08,2) {0.16};
\draw[fill=white, draw=black] (axis cs:0.05781375,0.7335) rectangle (axis cs:0.0736741666666667,0.84);
\draw[fill=white, draw=black] (axis cs:0.2858736875,1.7335) rectangle (axis cs:0.35039265625,1.84);
\node[align=center, font=\color{white}]
at (axis cs:0.255,2) {0.19};
\draw[fill=white, draw=black] (axis cs:0.1395790625,0.7335) rectangle (axis cs:0.206120104166667,0.84);
\node[align=center, font=\color{white}]
at (axis cs:0.14,1) {0.13};
\draw[fill=white, draw=black] (axis cs:0.41379159375,1.7335) rectangle (axis cs:0.48364428125,1.84);
\node[align=center, font=\color{white}]
at (axis cs:0.417,2) {0.13};
\node[align=center, font=\color{white}]
at (axis cs:0.608,2) {0.25};
\draw[fill=white, draw=black] (axis cs:0.303681707291667,0.7335) rectangle (axis cs:0.354690488541667,0.84);
\node[align=center, font=\color{white}]
at (axis cs:0.287,1) {0.14};
\draw[fill=white, draw=black] (axis cs:0.947556815625,1.7335) rectangle (axis cs:0.978802334375,1.84);
\node[align=center, font=\color{white}]
at (axis cs:0.871,2) {0.22};
\draw[fill=white, draw=black] (axis cs:0.3715483,0.7335) rectangle (axis cs:0.4351433,0.84);
\node[align=center]
at (axis cs:0.395,1) {0.08};
\draw[fill=white, draw=black] (axis cs:1.06334656875,1.7335) rectangle (axis cs:1.131382196875,1.84);
\node[align=center]
at (axis cs:1.055,2) {0.15};
\node[right, align=left]
at (axis cs:1.131,2) { 1.1};
\node[right, align=left]
at (axis cs:0.435,1) { 0.44};
\end{axis}
\end{tikzpicture}%

%% file: figures/scalability_gpu.tikz
%
\definecolor{mycolor1}{rgb}{0.00000,0.44700,0.74100}%
\definecolor{mycolor2}{rgb}{0.85000,0.32500,0.09800}%
\definecolor{mycolor3}{rgb}{0.92900,0.69400,0.12500}%
\definecolor{mycolor4}{rgb}{0.49400,0.18400,0.55600}%
\begin{tikzpicture}

\begin{axis}[%
unbounded coords=jump,
xmode=log,
xmin=2.82842712474619,
xmax=32768,
xtick={    4,     8,    16,    32,    64,   128,   256,   512,  1024,  2048,  4096,  8192, 16384, 32768},
xticklabels={    4,     8,    16,    32,    64,   128,   256,   512,  1024,  2048,  4096,  8192, \tiny 16384, \tiny 32768},
xlabel style={font=\color{white!15!black}},
xlabel={Number of GPUs},
ymode=log,
ymin=0.5,
ymax=1024,
ytick={   1,    2,    4,    8,   16,   32,   64,  128,  256,  512, 1024},
yticklabels={   1,    2,    4,    8,   16,   32,   64,  128,  256,  512, 1024},
ylabel style={font=\color{white!15!black}},
ylabel={Speed-up},
axis background/.style={fill=white},
xmajorgrids,
ymajorgrids,
legend style={legend cell align=left, align=left, draw=white!15!black, at={(0.5,1.03)}, anchor=south, legend columns=4, transpose legend, /tikz/every even column/.append style={column sep=5mm}, draw=none},
width=\figurewidth,
height=\figureheight,
]
\addplot [color=black, densely dashed, forget plot]
  table[row sep=crcr]{%
4	1\\
4096	1024\\
};

\addplot [color=mycolor1, line width=1pt, mark=+, mark size=2pt, mark options={solid, mycolor1}, forget plot]
  table[row sep=crcr]{%
6	0.870817973516165\\
12	1.69996130613378\\
24	3.27694457949618\\
48	5.78189334314224\\
96	9.24894710957223\\
192	13.0661977330227\\
};
\addplot [color=mycolor1, line width=1pt]
  table[row sep=crcr]{%
0.5	0.5\\
};
\addlegendentry{20M guiding centers per species, $32\times256\times128$ cells}

\addplot [color=mycolor2, densely dotted, line width=1pt, mark=+, mark size=2pt, mark options={solid, mycolor2}, forget plot]
  table[row sep=crcr]{%
48	6.76447534902506\\
96	12.3859784954237\\
192	22.7417510903428\\
384	36.7844238731521\\
768	47.0450068762231\\
1536	49.9613622220883\\
};
\addplot [color=mycolor2, densely dotted, line width=1pt]
  table[row sep=crcr]{%
0.5	0.5\\
};
\addlegendentry{160M guiding centers per species, $64\times512\times256$ cells}

\addplot [color=mycolor3, densely dashed, line width=1pt, mark=+, mark size=2pt, mark options={solid, mycolor3}, forget plot]
  table[row sep=crcr]{%
384	45.221705870696\\
768	72.916735159596\\
1536	124.229248404704\\
3072	145.565334039795\\
6144	163.602191279005\\
12288	165.6537342562\\
};
\addplot [color=mycolor3, densely dashed, line width=1pt]
  table[row sep=crcr]{%
0.5	0.5\\
};
\addlegendentry{1.3G guiding centers per species, $128\times1024\times512$ cells}

\addplot [color=mycolor4, dashdotted, line width=1pt, mark=+, mark size=2pt, mark options={solid, mycolor4}, forget plot]
  table[row sep=crcr]{%
3072	232.109049526848\\
6144	354.850769873332\\
12288	402.704673086968\\
24576	395.202786299591\\
49152	nan\\
98304	nan\\
};
\addplot [color=mycolor4, dashdotted, line width=1pt]
  table[row sep=crcr]{%
0.5	0.5\\
};
\addlegendentry{10G guiding centers per species, $256\times2048\times1024$ cells}

\addplot [color=mycolor1, line width=1pt, mark=o, mark size=1.5pt, mark options={solid, mycolor1}, forget plot]
  table[row sep=crcr]{%
1	nan\\
2	nan\\
4	1\\
8	1.67178996577012\\
16	2.91332560910298\\
32	3.38351402268662\\
};

\addplot [color=mycolor2, densely dotted, line width=1pt, mark=o, mark size=1.5pt, mark options={solid, mycolor2}, forget plot]
  table[row sep=crcr]{%
8	nan\\
16	nan\\
32	5.29895429252563\\
64	8.38205253071556\\
128	18.2203981177673\\
256	29.3487249858794\\
};
\addplot [color=mycolor3, densely dashed, line width=1pt, mark=o, mark size=1.5pt, mark options={solid, mycolor3}, forget plot]
  table[row sep=crcr]{%
64	nan\\
128	nan\\
256	24.4398898351561\\
512	34.1577038673589\\
1024	47.424834413808\\
2048	43.4755974939762\\
};
\addplot [color=mycolor4, dashdotted, line width=1pt, mark=o, mark size=1.5pt, mark options={solid, mycolor4}, forget plot]
  table[row sep=crcr]{%
512	nan\\
1024	nan\\
2048	109.025583079651\\
4096	nan\\
8192	nan\\
16384	nan\\
};

\addplot [color=black, densely dashed]
  table[row sep=crcr]{%
0.5	0.5\\
};
\addlegendentry{Ideal scaling}

\addplot [color=gray, line width=1pt, only marks, mark=+, mark size=2pt, mark options={solid}]
  table[row sep=crcr]{%
0.5	0.5\\
};
\addlegendentry{Summit}

\addplot [color=gray, dashed, line width=1pt, only marks, mark=o, mark size=1.5pt, mark options={solid}]
  table[row sep=crcr]{%
0.5	0.5\\
};
\addlegendentry{Piz Daint}

\end{axis}
\end{tikzpicture}%

%% file: ORB5.bbl
\begin{thebibliography}{10}

\bibitem{Garbet2010}
X.~Garbet, Y.~Idomura, L.~Villard, and T.H. Watanabe.
\newblock {Gyrokinetic simulations of turbulent transport}.
\newblock {\em Nucl. Fusion}, 50(4):043002, apr 2010.

\bibitem{Dannert2014}
T.~Dannert, A.~Marek, and M.~Rampp.
\newblock {Porting Large HPC Applications to GPU Clusters: The Codes GENE and
  VERTEX}.
\newblock {\em Adv. Parallel Comput.}, 25:305--314, 2014.

\bibitem{Sfiligoi2018}
I~Sfiligoi, J~Candy, and M~Kostuk.
\newblock {CGYRO Performance on Power9 CPUs and Volta GPUs}.
\newblock In {\em Lect. Notes Comput. Sci. (including Subser. Lect. Notes
  Artif. Intell. Lect. Notes Bioinformatics)}, volume 11203 LNCS, pages
  365--372. Springer, Cham, jun 2018.

\bibitem{Meng2013}
X.~Meng, X.~Zhu, P.~Wang, Y.~Zhao, X.~Liu, B.~Zhang, Y.~Xiao, W.~Zhang, and
  Z.~Lin.
\newblock Heterogeneous programming and optimization of gyrokinetic toroidal
  code and large-scale performance test on th-1a.
\newblock In Julian~Martin Kunkel, Thomas Ludwig, and Hans~Werner Meuer,
  editors, {\em Supercomputing}, pages 81--96, Berlin, Heidelberg, 2013.
  Springer Berlin Heidelberg.

\bibitem{Madduri2011}
K.~Madduri, K.~Ibrahim, S.~Williams, E.-J. Im, S.~Ethier, J.~Shalf, and
  L.~Oliker.
\newblock {Gyrokinetic toroidal simulations on leading multi- and manycore HPC
  systems}.
\newblock In {\em Proc. 2011 Int. Conf. High Perform. Comput. Networking,
  Storage Anal. - SC '11}, page~1, New York, New York, USA, 2011. ACM Press.

\bibitem{Azevedo2017}
E.~D'Azevedo, S.~Abbott, T.~Koskela, P.~Worley, S.-H. Ku, S.~Ethier, E.~Yoon,
  M.~S. Shephard, R.~Hager, J.~Lang, et~al.
\newblock The fusion code xgc: Enabling kinetic study of multiscale edge
  turbulent transport in iter.
\newblock In {\em Exascale Scientific Applications}, pages 529--552. Chapman
  and Hall/CRC, 2017.

\bibitem{Abbott2016}
S.~Abbott and E.~D'Azevedo.
\newblock Physics based optimization of particle-in-cell simulations on gpus.
\newblock In {\em APS Meeting Abstracts}, 2016.

\bibitem{Tang2017}
W.~Tang and Z.~Lin.
\newblock Global gyrokinetic particle-in-cell simulation.
\newblock In {\em Exascale Scientific Applications}, pages 507--528. Chapman
  and Hall/CRC, 2017.

\bibitem{Tran1999}
T.-M. Tran, K.~Appert, M.~Fivaz, G.~Jost, J.~Vaclavik, and L.~Villard.
\newblock {Global gyrokinetic simulation of ion-temperature-gradient-driven
  instabilities using particles}.
\newblock In J.~W. Connor, E.~Sindoni, and J.~Vaclavik, editors, {\em Th. of
  Fusion Plasmas}, volume~18, page~45. Societ{\`{a}} Italiana di Fisica,
  Bologna, 1999.

\bibitem{Jolliet2007}
S.~Jolliet, A.~Bottino, P.~Angelino, R.~Hatzky, T.-M. Tran, B.F. Mcmillan,
  O.~Sauter, K.~Appert, Y.~Idomura, and L.~Villard.
\newblock {A global collisionless PIC code in magnetic coordinates}.
\newblock {\em Comput. Phys. Commun.}, 177(5):409--425, sep 2007.

\bibitem{Bottino2011}
A.~Bottino, T.~Vernay, B.~Scott, S.~Brunner, R.~Hatzky, S.~Jolliet, B.F.
  McMillan, T.-M. Tran, and L.~Villard.
\newblock {Global simulations of tokamak microturbulence: Finite-$\beta$
  effects and collisions}.
\newblock {\em Plasma Phys. Control. Fusion}, 53(12):124027, dec 2011.

\bibitem{Lanti2019}
E.~Lanti, N.~Ohana, N.~Tronko, T.~Hayward-Schneider, A.~Bottino, B.F. McMillan,
  A.~Mishchenko, A.~Scheinberg, A.~Biancalani, P.~Angelino, S.~Brunner,
  J.~Dominski, P.~Donnel, C.~Gheller, R.~Hatzky, A.~Jocksch, S.~Jolliet, Z.X.
  Lu, J.P. {Martin Collar}, I.~Novikau, E.~Sonnendr{\"{u}}cker, T.~Vernay, and
  L.~Villard.
\newblock {Orb5: A global electromagnetic gyrokinetic code using the PIC
  approach in toroidal geometry}.
\newblock {\em Comput. Phys. Commun.}, page 107072, nov 2019.

\bibitem{Kim2000}
C.C. Kim and S.E. Parker.
\newblock {Massively Parallel Three-Dimensional Toroidal Gyrokinetic Flux-Tube
  Turbulence Simulation}.
\newblock {\em J. Comput. Phys.}, 161(2):589--604, jul 2000.

\bibitem{Hatzky2006}
R.~Hatzky.
\newblock {Domain cloning for a particle-in-cell (PIC) code on a cluster of
  symmetric-multiprocessor (SMP) computers}.
\newblock {\em Parallel Comput.}, 32(4):325--330, apr 2006.

\bibitem{git}
{Git web homepage}.
\newblock \url{https://git-scm.com}.

\bibitem{jenkins}
{Jenkins web homepage}.
\newblock \url{https://jenkins.io}.

\bibitem{Scheinberg2019}
A.~Scheinberg, S.~Ethier, C.-S. Chang, G.~Chen, R.~Bird, S.~Slattery, and
  P.~Worley.
\newblock {Kokkos and Fortran in the Exascale Computing Project plasma physics
  code XGC}.
\newblock In {\em Proc. 2019 Int. Conf. High Perform. Comput. Networking,
  Storage Anal. - SC '19}, page~3. ACM Press, 2019.

\bibitem{Beckingsale2019}
D.~A. Beckingsale, T.~R.~W. Scogland, J.~Burmark, R.~Hornung, H.~Jones,
  W.~Killian, A.~J. Kunen, O.~Pearce, P.~Robinson, and B.~S. Ryujin.
\newblock {RAJA: Portable Performance for Large-Scale Scientific Applications}.
\newblock In {\em 2019 IEEE/ACM International Workshop on Performance,
  Portability and Productivity in HPC (P3HPC)}, pages 71--81. IEEE, nov 2019.

\bibitem{Keryell2019}
R.~Keryell.
\newblock {SYCL: A Single-Source C++ Standard for Heterogeneous Computing}.
\newblock In {\em 2019 IEEE/ACM International Workshop on Performance,
  Portability and Productivity in HPC (P3HPC)}, 2019.

\bibitem{Sugama2000}
H.~Sugama.
\newblock {Gyrokinetic field theory}.
\newblock {\em Phys. Plasmas}, 7(2):466--480, feb 2000.

\bibitem{Tronko2017}
N.~Tronko, A.~Bottino, C.~Chandre, and E.~Sonnendr{\"{u}}cker.
\newblock {Hierarchy of second order gyrokinetic Hamiltonian models for
  particle-in-cell codes}.
\newblock {\em Plasma Phys. Control. Fusion}, 59(6):064008, jun 2017.

\bibitem{Lutjens1996}
H.~L{\"{u}}tjens, A.~Bondeson, and O.~Sauter.
\newblock {The CHEASE code for toroidal MHD equilibria}.
\newblock {\em Comput. Phys. Commun.}, 97(3):219--260, sep 1996.

\bibitem{Dominski2017}
J.~Dominski, B.F. McMillan, S.~Brunner, G.~Merlo, T.-M. Tran, and L.~Villard.
\newblock {An arbitrary wavelength solver for global gyrokinetic simulations.
  Application to the study of fine radial structures on microturbulence due to
  non-adiabatic passing electron dynamics}.
\newblock {\em Phys. Plasmas}, 24(2):022308, feb 2017.

\bibitem{Vernay2010}
T.~Vernay, S.~Brunner, L.~Villard, B.F. McMillan, S.~Jolliet, T.-M. Tran,
  A.~Bottino, and J.P. Graves.
\newblock {Neoclassical equilibria as starting point for global gyrokinetic
  microturbulence simulations}.
\newblock {\em Phys. Plasmas}, 17(12):122301, dec 2010.

\bibitem{McMillan2008}
B.F. McMillan, S.~Jolliet, T.-M. Tran, L.~Villard, A.~Bottino, and P.~Angelino.
\newblock {Long global gyrokinetic simulations: Source terms and particle noise
  control}.
\newblock {\em Phys. Plasmas}, 15(5):052308, may 2008.

\bibitem{McMillan2011}
B.F. McMillan, P.~Hill, A.~Bottino, S.~Jolliet, T.~Vernay, and L.~Villard.
\newblock {Interaction of large scale flow structures with gyrokinetic
  turbulence}.
\newblock {\em Phys. Plasmas}, 18(11):112503, nov 2011.

\bibitem{Jolliet2009}
S.~Jolliet.
\newblock {\em {Gyrokinetic Particle-In-Cell Global Simulations of
  Ion-Temperature-Gradient and Collisionless-Trapped- Electron-Mode Turbulence
  in Tokamaks}}.
\newblock PhD thesis, EPFL, Lausanne, 2009.

\bibitem{Mishchenko2017}
A.~Mishchenko, A.~Bottino, R.~Hatzky, E.~Sonnendr{\"{u}}cker, R.~Kleiber, and
  A.~K{\"{o}}nies.
\newblock {Mitigation of the cancellation problem in the gyrokinetic
  particle-in-cell simulations of global electromagnetic modes}.
\newblock {\em Phys. Plasmas}, 24(8):081206, aug 2017.

\bibitem{Mishchenko2019}
A.~Mishchenko, A.~Bottino, A.~Biancalani, R.~Hatzky, T.~Hayward-Schneider,
  N.~Ohana, E.~Lanti, S.~Brunner, L.~Villard, M.~Borchardt, R.~Kleiber, and
  A.~K{\"{o}}nies.
\newblock {Pullback scheme implementation in ORB5}.
\newblock {\em Comput. Phys. Commun.}, 238:194--202, may 2019.

\bibitem{Jocksch2017}
A.~Jocksch, N.~Ohana, E.~Lanti, A.~Scheinberg, S.~Brunner, C.~Gheller, and
  L.~Villard.
\newblock Prediction of the inter-node communication costs of a new gyrokinetic
  code with toroidal domain.
\newblock In R.~Wyrzykowski, J.~Dongarra, E.~Deelman, and K.~Karczewski,
  editors, {\em Parallel Processing and Applied Mathematics}, pages 370--380,
  Cham, 2018. Springer International Publishing.

\bibitem{Hariri2016}
F.~Hariri, T.-M. Tran, A.~Jocksch, E.~Lanti, J.~Progsch, P.~Messmer,
  S.~Brunner, C.~Gheller, and L.~Villard.
\newblock A portable platform for accelerated pic codes and its application to
  gpus using openacc.
\newblock {\em Computer Physics Communications}, 207:69 -- 82, 2016.

\bibitem{Ohana2016}
N.~Ohana, A.~Jocksch, E.~Lanti, T.-M. Tran, S.~Brunner, C.~Gheller, F.~Hariri,
  and L.~Villard.
\newblock {Towards the optimization of a gyrokinetic Particle-In-Cell (PIC)
  code on large-scale hybrid architectures}.
\newblock In {\em J. Phys. Conf. Ser.}, volume 775, page 012010. IOP
  Publishing, nov 2016.

\bibitem{OpenMP}
{OpenMP web homepage}.
\newblock \url{https://www.openmp.org/}.

\bibitem{OpenACC}
{OpenACC web homepage}.
\newblock \url{https://www.openacc.org/}.

\bibitem{Jocksch2015}
A.~Jocksch, F.~Hariri, T.-M. Tran, S.~Brunner, C.~Gheller, and L.~Villard.
\newblock A bucket sort algorithm for the particle-in-cell method on manycore
  architectures.
\newblock In R.~Wyrzykowski, E.~Deelman, J.~Dongarra, K.~Karczewski,
  J.~Kitowski, and K.~Wiatr, editors, {\em Parallel Processing and Applied
  Mathematics}, pages 43--52, Cham, 2016. Springer International Publishing.

\end{thebibliography}
